\documentclass[usenatbib,usegraphicx]{mn2e}
\usepackage[letterpaper,margin=0.65in]{geometry}
\usepackage{amsmath,amssymb,amsfonts} 
\usepackage{graphics}                 
\usepackage{color}                    
\usepackage{booktabs}
\usepackage{multirow}
\usepackage{url}

\newcommand{\ie}{{i.e., }}

\title{On the Clustering of Compact Galaxy Pairs in Dark Matter Haloes}

\author[Y. Wang et al.]{Y.~Wang$^1$, R.~J.~Brunner$^1$\\
$^1$Department of Astronomy, University of Illinois, 1002 W. Green St., Urbana, IL 61801, USA}

\begin{document}
\date{Draft: \today}

\maketitle
\begin{abstract}
{
{We analyze the clustering of photometrically selected galaxy pairs by using the halo-occupation distribution (HOD) model. We measure the angular two-point auto-correlation function, $\omega(\theta)$, for galaxies and galaxy pairs in three volume-limited samples and develop an HOD to model their clustering. Our results are successfully fit by these HOD models, and we see the separation of ``1-halo" and "2-halo" clustering terms for both single galaxies and galaxy pairs. Our clustering measurements and HOD model fits for the single galaxy samples are consistent with previous results. We find that the galaxy pairs generally have larger clustering amplitudes than single galaxies, and the quantities computed during the HOD fitting, e.g., effective halo mass, $M_{eff}$, and linear bias, $b_{g}$, are also larger for galaxy pairs. We find that the central fractions for galaxy pairs are significantly higher than single galaxies, which confirms that galaxy pairs are formed at the center of more massive dark matter haloes. We also model the clustering dependence of the galaxy pair correlation function on redshift, galaxy type, and luminosity. We find early-early pairs (bright galaxy pairs) cluster more strongly than late-late pairs (dim galaxy pairs), and that the clustering does not depend on the luminosity contrast between the two galaxies in the compact group.}
}
\end{abstract}


\section{Introduction}\label{intro}
Studying the clustering of galaxies provides insights into both galaxy formation and evolution, especially when analyzed as a function of the intrinsic properties of a galaxy such as mass, size, morphology, gas content, star-formation rate, nuclear activities, characteristic velocity and metallicity~{\citep{Kauf04, Yang12, WatsonSFR}}. Traditional clustering analyses can connect the visible galaxies to the unseen dark matter haloes, thereby providing insight into the assembly history of galaxies. The standard picture in an $\Lambda$CDM universe is that galaxies are formed from gas that cools and condenses at the centers of dark matter haloes~\citep{White1978}. As smaller haloes fall into more massive ones, their central galaxies become satellites of the new, larger hosts, sometimes merging to form new central galaxies in the core of the resultant dark matter halo. In this model, each halo will contain a central, dominant galaxy at the bottom of the gravitational potential well, surrounded by other, smaller satellite galaxies that orbit around the dominant galaxy.

{While successful in predicting the clustering behavior of single galaxies~\citep{Zehavi04}, galaxy clustering studies typically have not analyzed groups of galaxies, yet we know that many galaxies live in dense environments. {While \citet{Berlind} did study the multiplicity functions of groups that are selected from SDSS; in this paper, however, we extend our previous SDSS clustering results~\citep{Wang13} to the study of the clustering of galaxy groups.} As a result, we first construct a catalog of photometrically selected compact galaxy groups, and second use this new catalog to measure and analyze the clustering behavior of small galaxy groups in the same framework traditionally applied to isolated galaxies. Compact groups contain a small number of galaxies with compact angular configurations~\citep{Hickson} where galaxy-galaxy interactions are the dominant processes. On the other hand, rich groups or galaxy clusters evolve under both galaxy-galaxy interactions~\citep{Toomre, Farou} and galaxy-environment interactions~\citep{Gunn, Nulsen}.}

As a result, the abundance of compact groups, their spatial distribution, and their intrinsic properties are entwined with their halo merger history, which is closely related to the underlying cosmology. For example, the evolution of merger rates is sensitive to the cosmic matter density, while the mass distribution of merging objects depends on the linear power spectrum of the initial density of fluctuations~\citep{Lacey}. Thus, the properties of compact groups can be used to constrain cosmological parameters. On the other hand, the physical processes driving galaxy evolution have a strong effect on satellite galaxies. For example, galaxy colors are affected by the stripping of gas within the galaxies during interaction events, which can suppress star formation; while the galaxy structures and morphologies can be modified by the gravitational and hydrodynamical interactions between galaxies.

Modeling galaxy clustering is complicated by the fact that the gravitational framework is supplied by the invisible dark matter, thus we must have a means to connect the visible tracers of the gravitational field to the invisible dark matter structure. Currently, there are two methods that have been widely used to provide this connection. First is the use of a semi-analytic model (SAM) that employs a simplified representation of the relevant astrophysics to follow galaxy growth within the evolving dark matter halo population. The SAM can be used to predict the detailed properties of both central and satellite galaxies~\citep{White1991, Springel, Kang} for comparison with an observed galaxy population. {The second approach is the semi-empirical technique that provides an insightful description of the relations between the galaxies and their host dark matter halos via the observed galaxy properties. There are several different semi-empirical techniques in use, including abundance matching~\citep{Conroy, HearinAbM} and age matching~\citep{HearinAgM}; in this paper, however, we use the Halo Occupation Distribution~\citep[hereafter HOD;][]{Jing, Peacock, Zehavi04, Zheng05}.} An HOD model quantifies the central and satellite galaxy populations of dark matter haloes as a function of the host halo mass by optimizing the model fit to the measured galaxy correlation function from large surveys.

Since galaxies live in a variety of different environments, the measurement of their clustering properties as a function of scale has displayed a variance since they were initially measured~\citep{Peebles}. Initially, this difference in clustering was described by fitting a simple power-law function to the two-point galaxy correlation functions on small scales~\citep{Peebles}. However, more recent large area surveys have measured the clustering pattern accurately enough to detect deviations from the traditional power-law model~\citep[e.g.,][]{Zehavi04,Watson}. This deviation provides an important insight into the growth of large-scale structure; and, the deviations have been more accurately interpreted in terms of dark matter haloes, which link the small scale clustering to the underlying dark matter haloes. A well defined HOD, along with accurate correlation function measurements, can be used to statistically describe how galaxies populate dark matter halos as a function of halo mass. 

With the enormous data being provided by large surveys, such as the recently completed Sloan Digital Sky Survey~\citep[SDSS;][]{York}, this type of analysis can also now be extended to the study of galaxy groups. To date, there have been two primary approaches used to extract galaxy groups from the SDSS. First,~\cite{Yang07} selected galaxy groups from the SDSS spectroscopic data, which has a limiting {apparent} magnitude of $r \sim 17.7$, finding significant disagreement with a mock SAM catalog previously created by~\citet{Croton}. A second approach was developed by~\cite{Wang12}, {see also~\cite{Tala} and~\cite*{Talb}}, who combined the spectroscopic and photometric SDSS data to identify a spectroscopic central galaxy with photometric satellites, which allowed them to study the luminosity and mass functions of the satellite galaxies.

\begin{figure}
{\resizebox{9 cm}{!}{\includegraphics{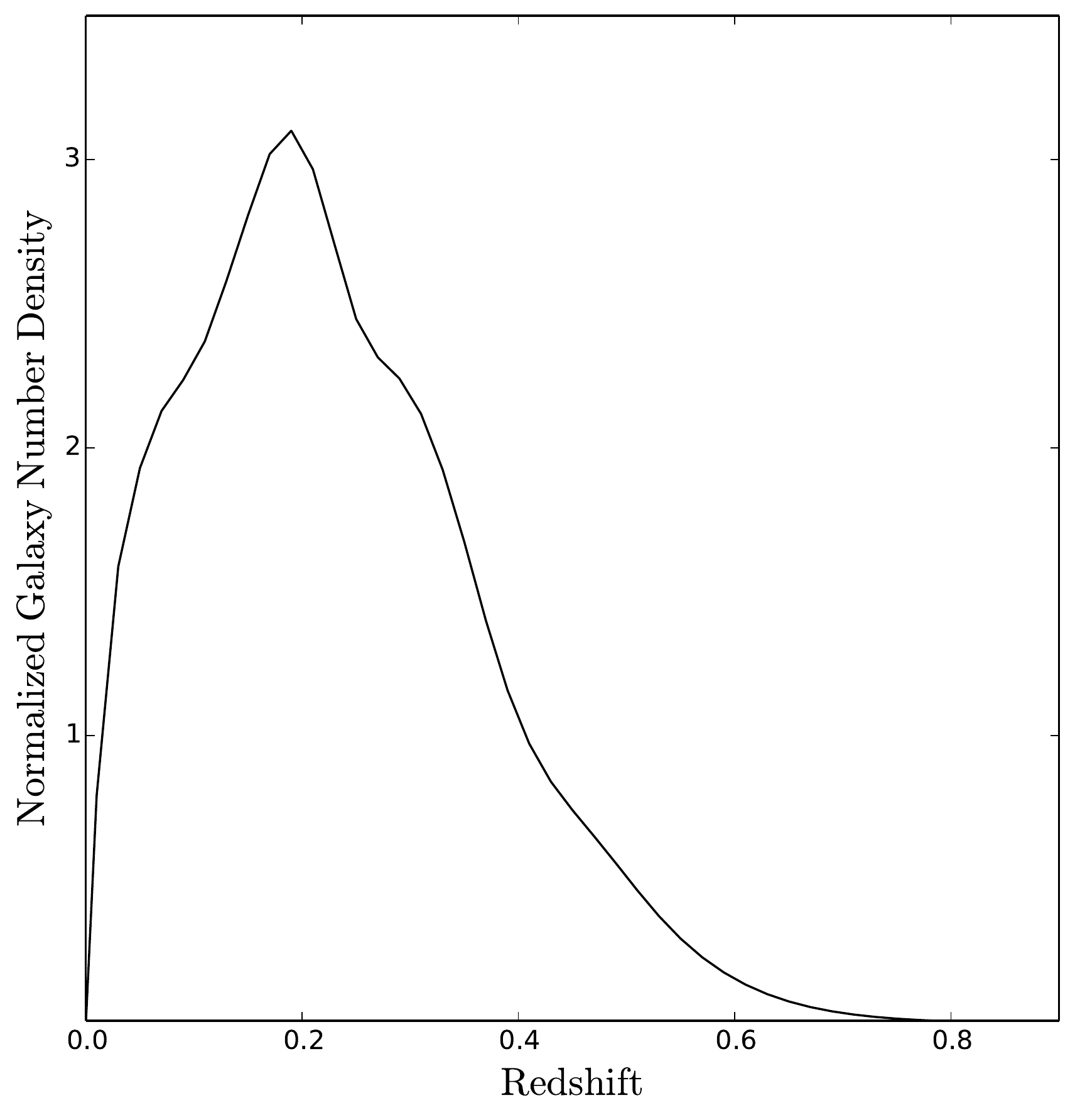}}}
\caption{The normalized galaxy redshift distribution for the SDSS DR7 galaxies in the Wang et al. (2013) clean galaxy catalog restricted to the range $18 < \textrm{r} < 21$. We assume a Gaussian photometric redshift probability density function for each galaxy with the Gaussian $\mu$ and $\sigma$ given by the photometric redshift and its associated error. The individual photometric redshift PDFs are stacked to construct this smooth figure. The mode of the redshift distribution lies at $z = 0.176$ and the median redshift is $<\textrm{z}> = 0.24$ with a dispersion of $0.16$. \label{NumDen}}
\end{figure}

In this paper, we extend the previous galaxy clustering analyses to galaxy groups. We start with the clean galaxy catalog we generated~\citep{Wang13} from the photometric data released in the seventh data release (DR7) of the SDSS~\citep{Aba09}.  We first select compact galaxy groups, following ~\cite{Hickson} to obtain a flux-limited compact galaxy group catalog. Since this a photometric only data set, we do not have spectroscopic redshifts to remove contamination from line-of-sight galaxy interlopers. While we would prefer to remove these galaxies, the standard techniques of bootstrap background correction~\citep{Lorri} or quantifying the galaxy cluster probabilistic membership via photometric redshifts~\citep{Brunner00} do not provide a reliable method to identify and remove interloping galaxies at the low redshifts of the galaxies in our photometric sample as shown in Figure~\ref{NumDen}. However, the correlation functions we will measure for the galaxy groups are much stronger than the correlation functions we measure for single galaxies. In addition, the group correlation functions still show deviations from a power-law model, thus we will still be able to connect the distribution of compact galaxy groups with their parent dark matter halos by developing a compact group HOD.

We first describe the clean galaxy sample and both the luminosity-limited and volume-limited sub-samples in Section~\ref{data}. Next, in Section~\ref{groups} we introduce the selection criteria for the angular-compact galaxy groups and discuss the basic properties of our group catalogs. In Section~\ref{cluster}, we measure the correlation functions of the galaxy groups and quantify their dependence on group richness and galaxy type. We present our new HOD model in Section~\ref{hod}, where we first apply a basic HOD model as shown in~\cite{Ross} to the isolated galaxies selected from the SDSS DR7, and second we apply our new HOD model to our galaxy pair catalog, and to our galaxy pair catalog split by galaxy type. Finally, we present the modeling result in Section~\ref{hod_singlePairs}, and conclude with a discussion of our results in Section~\ref{results}.

\section{Data}\label{data}

The first phase of the SDSS was a photometric and spectroscopic survey designed to map one-fourth of the entire sky to produce a large data set to analyze large-scale structure and to study the underlying cosmic evolution. The survey was conducted by the Astrophysical Research Consortium at the Apache Point Observatory in New Mexico, and provided photometric observations in five bands: $u,\ g,\ r,\ i,$ and $z$~\citep{Fuku}. The last data release from the first phase was DR7, which was released in November 2008, and included approximately 10$^8$ photometric galaxies to a 5$\sigma$ detection limit of r $\sim$ 23.1 covering approximately 10$^4$ deg$^2$ of the sky~\citep{Aba09}.

\begin{figure}
{{\includegraphics[width=0.48\textwidth]{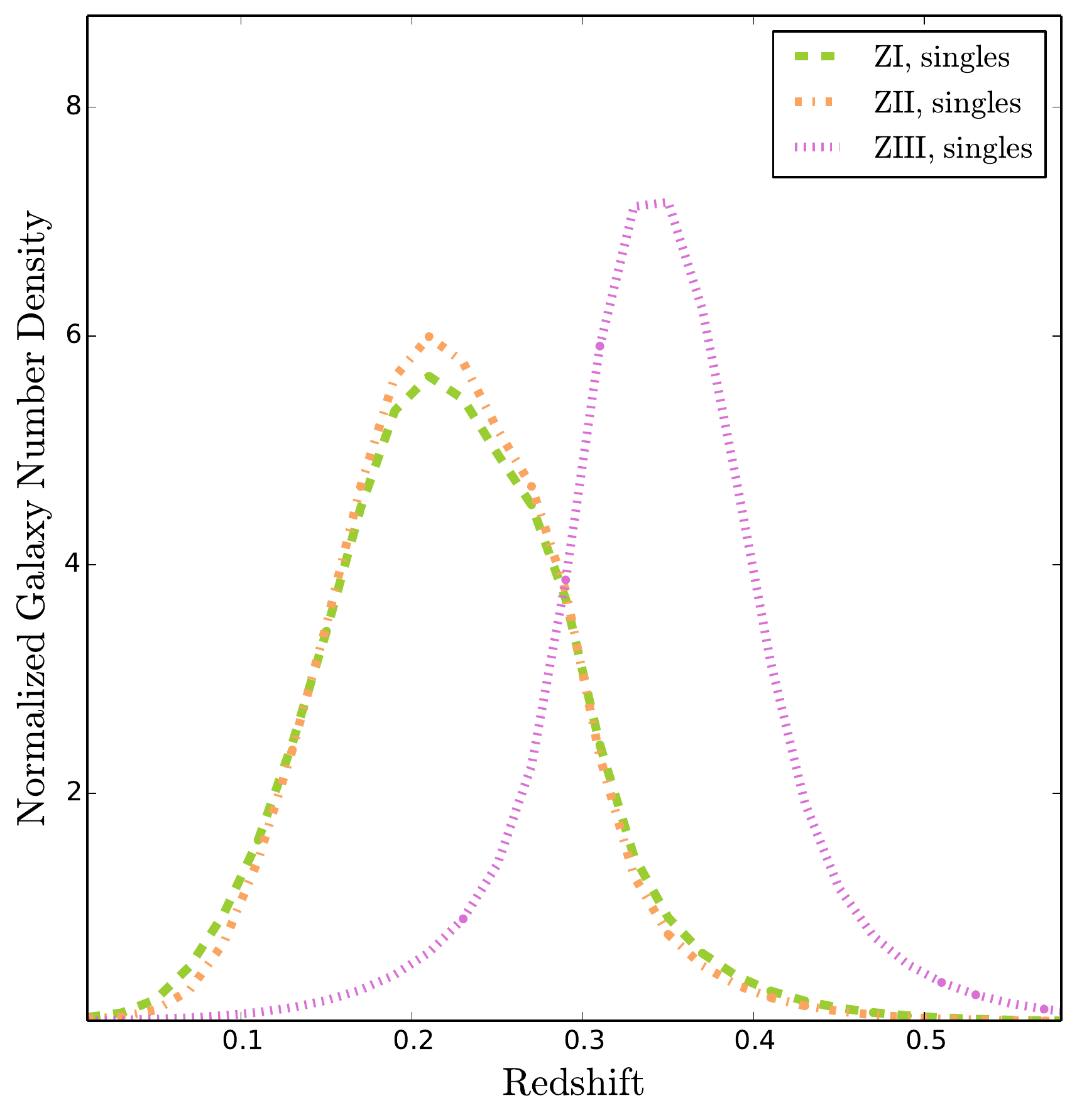}}}
\caption{The normalized galaxy redshift distribution for the three volume-limited samples: ZI (dashed), ZII (dash-dotted) and ZIII (dotted). The individual distributions are computed in the same manner as Figure~\ref{NumDen}.\label{nzVol}}
\end{figure}

As a starting point, we use our previously constructed clean galaxy sample from the SDSS DR7 (Wang et al. 2013). To construct this catalog, we analyzed a number of different SDSS flags to {optimally minimize the impact of galaxy image deblending and we} tested the best magnitude range to optimally select a complete galaxy catalog by comparing source detections from the single-image and stacked image Stripe 82 data. We also identified how to optimally mitigate the systematic effects of atmosphere seeing variations and Galactic extinction. A complete, detailed description of how the clean galaxy catalog was generated is outlined in Appendix A of~\citet{Wang13}.
 
In this work, we follow our previous guidelines and impose a seeing cut of $<1.3\arcsec$ and a reddening cut of $< 0.15$ mag. In the end, the initial catalog we use contains approximately 21 million galaxies with extinction-corrected $r$-band model magnitudes in the range $17 < r \leq 21$. The normalized galaxy redshift distribution for the entire galaxy catalog is presented in Figure~\ref{NumDen}. For this Figure, we assume each galaxy is represented in redshift space by a Gaussian photometric redshift probability density function with $\mu$ and $\sigma$ given by the photometric redshift value and error.

\subsection{Volume-Limited sub-samples}\label{volData}

To better understand the evolution of compact galaxy groups and their luminosity dependencies,  we generate three volume and luminosity-limited sub-samples according to the following prescriptions:\\
\begin{enumerate}
\renewcommand{\theenumi}{\Roman{enumi}:}
\item $0.0 < z \leq 0.3$, $M_{r}\ < -19.0$,
\item $0.0 < z \leq 0.3$, $M_{r}\ < -19.5$, and
\item $0.3 < z \leq 0.4$, $M_{r}\ < -19.5$.
\end{enumerate}

We denote these three sub-samples as ZI, ZII, and ZIII, and they contain approximately $4$ million, $2.2$ million, and $1.8$ million galaxies, respectively. These explicit redshift ranges were selected to result in a ratio of $1:1.2$ for the cosmic volumes of the $0.0 < z \leq 0.3$ and $0.3 < z \leq 0.4$ samples. These samples were $k$-corrected before construction, and the final galaxy redshift distributions are  shown in Figure~\ref{nzVol}, where the photometric redshift probability density functions were used in the same manner as for Figure~\ref{NumDen}.

\begin{figure*}
\begin{center}
{{\includegraphics[width=0.33\textwidth]{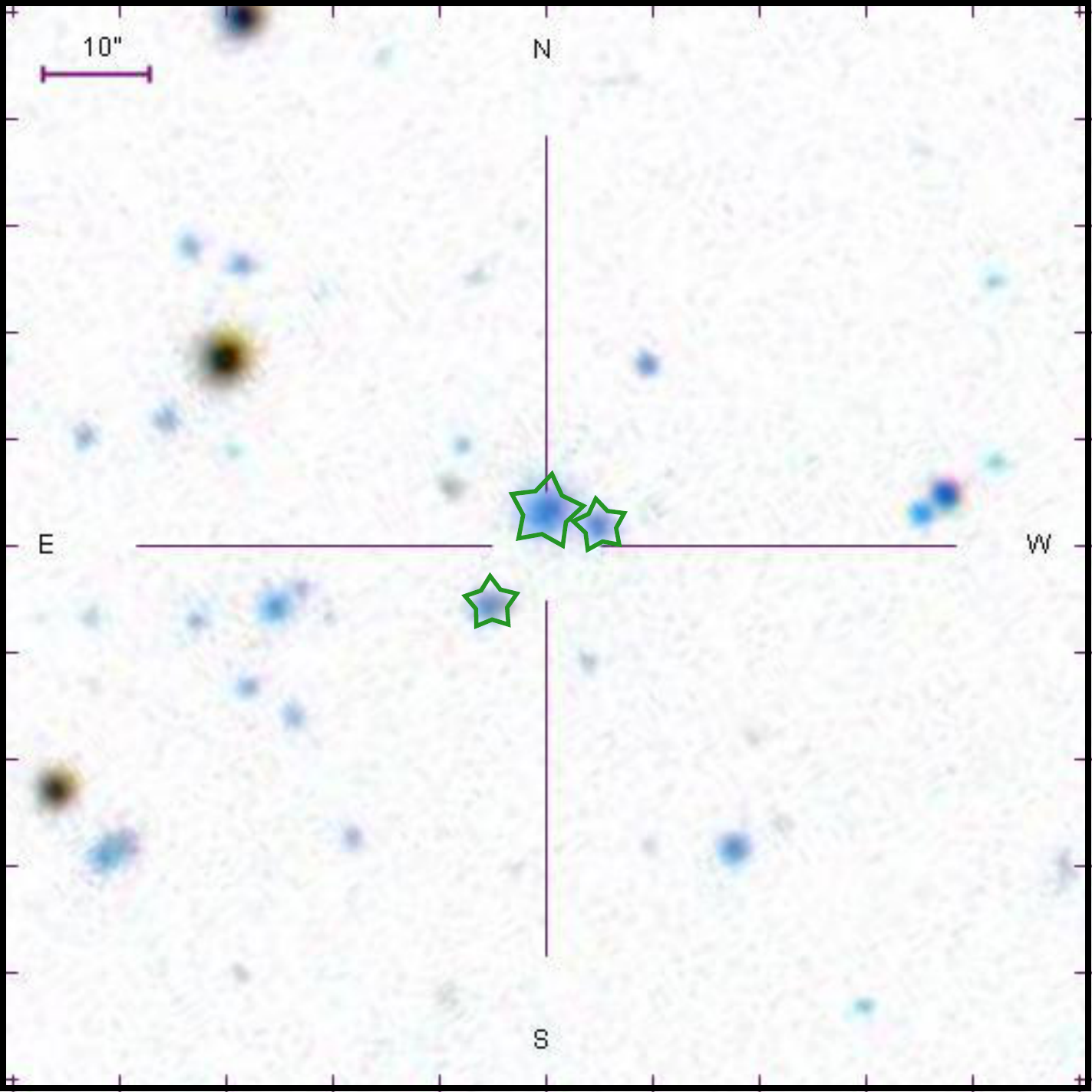}}}
{{\includegraphics[width=0.33\textwidth]{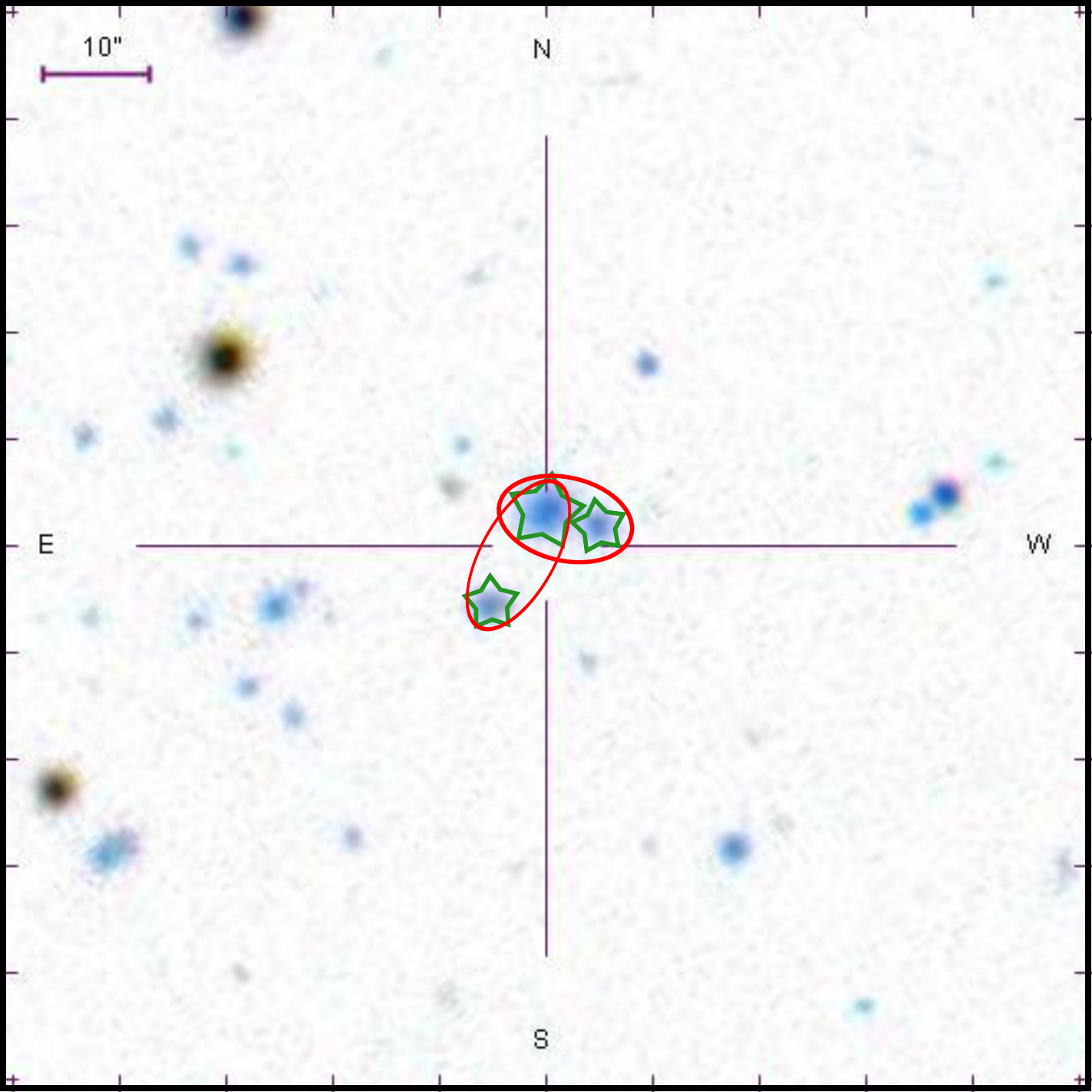}}}
{{\includegraphics[width=0.33\textwidth]{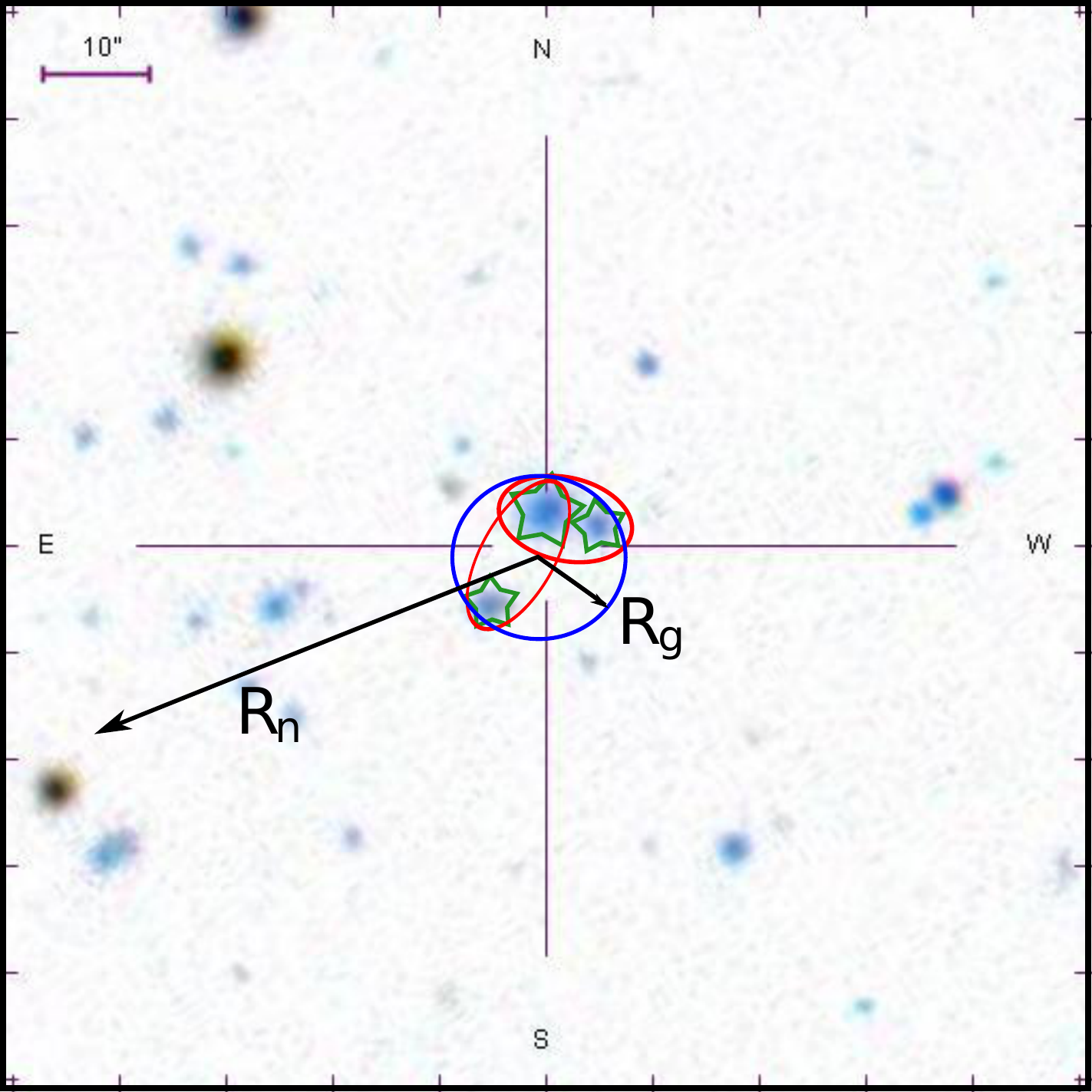}}}
\end{center}
\caption{A visual demonstration of the criteria we use to select isolated groups of galaxies. Left: The green star shapes are the galaxies in our ZI sample. Middle: The selected galaxies that are separated by between two and fifteen arcseconds are circled in red. Right: The compact, isolated groups are circled in blue with both $R_g$ and $R_n$ indicated by arrows for one group. This sample group is located at 115.331,+31.793.}
\label{sampleCatalog}
\end{figure*}

\section{Compact Galaxy Groups}\label{groups}

The primary focus of this paper is to analyze the clustering of compact galaxy groups. In this section, we present the method used to select compact galaxy groups from the SDSS DR7 clean galaxy catalog, which was discussed in Section~\ref{data}, and provide a brief analysis of this compact galaxy group catalog.

\subsection{Selection Criteria}

Based on the criteria presented by~\citet{Hickson}, we select isolated, compact groups of galaxies that satisfy the following four tests:
\begin{enumerate}
\renewcommand{\theenumi}{\arabic{enumi}:}
\item $2\arcsec <  \theta_{s} \leq 15\arcsec$,
\item $N = 2$, $3$, $4$, or $\geq 5$,
\item $\theta_{n} \geq 3 \theta_{g}$, and
\item $\mu_{g}  < 26.0$ mag arcsec$^{-2}$.
\end{enumerate}
This process, which is graphically demonstrated in Figure~\ref{sampleCatalog}, first selects close galaxy pairs, where $\theta_{s}$ is the separation between the two galaxies in the pair, which should be greater than $2\arcsec$ for the SDSS DR7 since the SDSS deblending algorithm rarely separate galaxies reliably at separations closer than this value~\citep{Infante}. $N$ is the number of galaxies within $3$ magnitudes of the brightest galaxy, and we create the final compact group catalog by compressing the selected galaxy pairs that have common members. $\theta_{g}$ is the angular radius of the isolated group, and $\theta_{n}$ is the maximum angular distance from the group center to the nearest neighbor galaxy. $\mu_{g}$ is the averaged surface brightness of these $N$ galaxies within the group angular radius.

\subsection{Compact Group Catalogues}

By applying the above selection criteria to the three volume-limited sub-samples, which were described in section~\ref{volData}, we construct catalogs consisting of isolated field galaxies ($N = 1$), galaxy pairs ($N = 2$), galaxy triplets ($N = 3$), and galaxy quads ($N = 4$). In Table~\ref{groupsRichnessTable} we present the number of compact galaxy groups as a function of group richness. The different catalogs all display a similar decrease with increasing group size. They also show a decrease in the total number of groups of a given size as the total number of galaxies in the input sample decreases (i.e., going from ZI to ZII/ZIII). As the final two sub-samples have similar number of galaxies (within 10\%), it is reassuring to see they have nearly identical group populations.

\begin{table}
\begin{center}
\caption{The number of compact galaxy groups in the SDSS DR7 for the three volume-limited samples presented in this paper.}
{\begin{tabular}{c  c  c  c  c  c  c  c}
\hline
$Catalog$ & $N = 1$ & $N = 2$ & $N = 3$ & $N = 4$ \\ \hline
ZI & 3,665,035 & 134,155 & 8,137 & 669  \\
ZII & 2,045,095 & 52,792 & 2,568 & 198 \\
ZIII & 1,699,581& 42,943 & 2,506 & 222  \\ \hline
\end{tabular}}
\label{groupsRichnessTable}
\end{center}
\end{table}

\subsection{Estimating Group Redshifts}\label{est_z}

Since we are identifying compact galaxy groups directly from photometric data, we do not have spectroscopic redshifts. While we could use photometric redshifts, their relatively large errors bars, especially at the low redshifts probed by the SDSS, make it difficult to quantify a group redshift. We, therefore, explore four different techniques for estimating the redshift of a compact galaxy group. Several of these techniques treat a photometric redshift in a similar manner as a spectroscopic redshift, while others treat the each photometric redshift estimate as a Gaussian probability density function (PDF), where the Gaussian mean and width are given by the estimated photometric redshift and error. We note that, if available, these approaches would likely be improved by using more accurate photometric redshift PDFs~\citep[e.g.,][]{TPZ,SOMZ}.

The first technique is to compute the average photometric redshift of all galaxies in the compact group. The second technique sums the individual galaxy photometric redshift PDFs, while the third technique multiplies the photometric redshift PDFs to generate the group redshift estimate. The fourth technique uses the estimated photometric redshifts to identify the brightest, in absolute magnitude as computed by using the  photometric redshifts, compact group galaxy. The photometric redshift of this brightest galaxy is used as a proxy for the group redshift. 

\begin{figure}
\begin{center}
{{\includegraphics[width=0.48\textwidth]{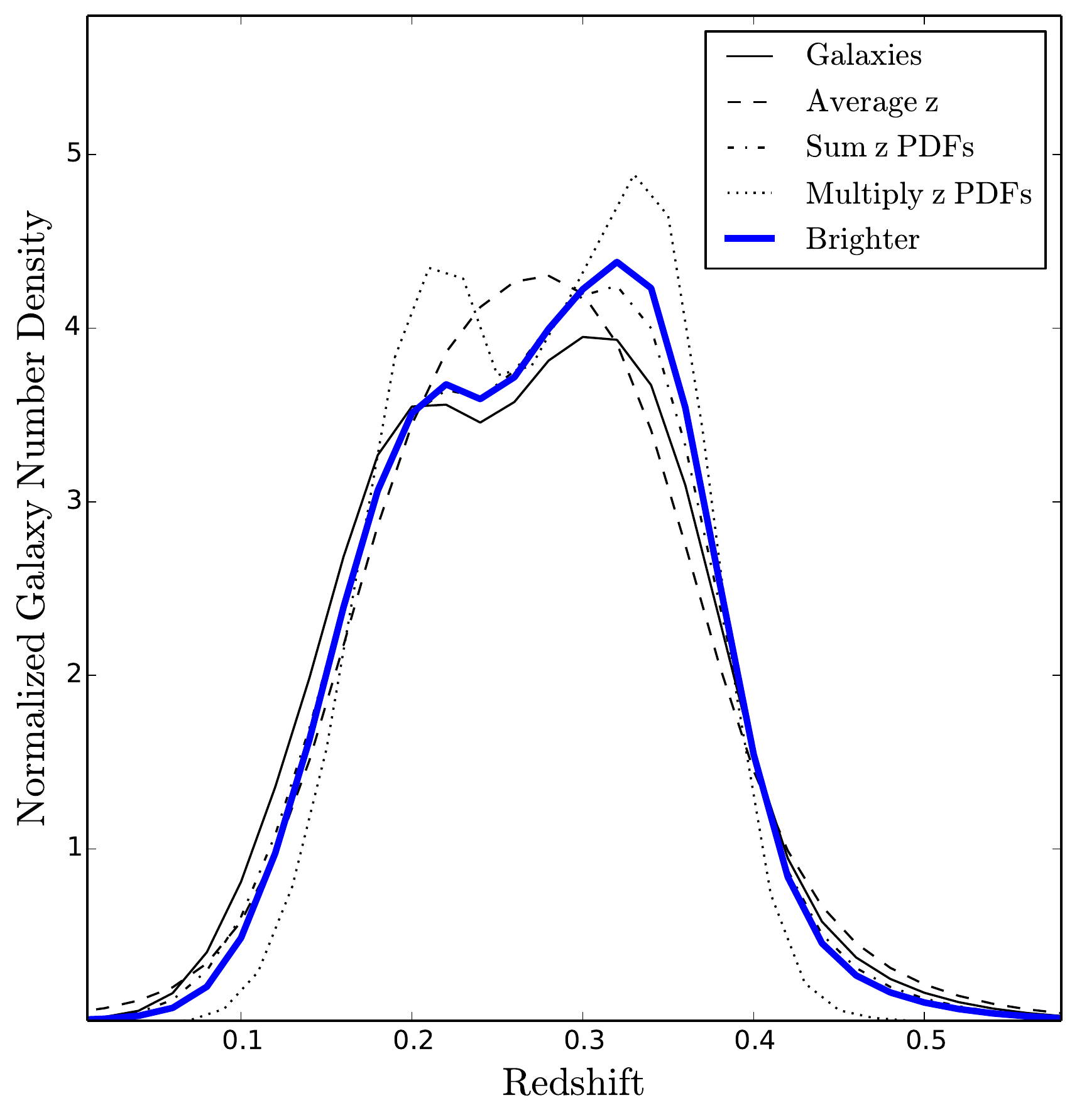}}}
\end{center}
\caption{A comparison, using only close galaxy pairs, of the four methods we developed to compute the photometric redshift of a compact galaxy group to the redshift distribution of the main galaxy sample. The method we use is highlighted in blue.
\label{nzgroupCompare}}
\end{figure}

We compare these four techniques for the galaxy pair sample over the full redshift range of the main volume-limited sample (0.0 $<$ z $\leq$ 0.4) to the main galaxy redshift distribution in Figure~\ref{nzgroupCompare}. We find that only one method, averaging the photometric redshifts, performs poorly, when compared to the main galaxy redshift distribution. Two techniques, the summation of the different PDFs and simply selecting the brightest group member's photometric redshift seem to perform the best. As a result, we choose the simplest technique, and use the photometric redshift of the brightest group member as a proxy for the group's photometric redshift (highlighted in blue). Given our chosen definition for group redshift, it is not surprising that the normalized redshift distributions for the pairs are nearly identical to the galaxy redshift distributions in each volume limited sub-sample, which are shown in Figure~\ref{nzVol}.

\section{Clustering Properties}\label{cluster}

To understand the clustering of compact galaxy groups, especially in comparison with normal galaxies, we will use the two-point correlation function. In this section, we present how we compute the correlation functions for different samples and briefly discuss the results.

\subsection{Methodology}

The two-point angular correlation function (TPACF) is a simple, yet effective technique for quantifying the clustering of a spatial point process. The TPACF measures the excess probability over random that one object will be located at a certain distance from another object. In our case, the object can be a galaxy, a galaxy pair, a galaxy triple, or a galaxy quad. Since the compact groups have already been preselected to be clustered (\ie they are already overdense), we expect their clustering signals to be enhanced relative to the full galaxy sample, which corresponds to them being more strongly correlated.

To speed up the bin counting process when estimating the correlation functions, we use our publicly available, fast two-point correlation code\footnote{\url{http://lcdm.astro.illinois.edu/code}} that implements a two-dimensional quad-tree structure to vastly reduce the computational time. More details about this code can be found in~\cite{Dol} and~\cite{Wang13}.

For the three samples described in Section~\ref{volData}, the correlation function is calculated between 0$^\circ$.01 and 10$^\circ$ with a logarithmic binning of twenty-five bins in total angle separation. It is processed by cutting into thirty-two sub-samples, which can be used to calculate jackknife errors. We find thirty-two sub-samples are sufficient enough to create a stable covariance matrix. To maintain a sufficient signal to noise ratio in the correlation measurements, we restrict the angular ranges for the correlation functions and the covariance matrices to be between 0$^\circ$.01 to 1$^\circ$.5, which provide nineteen bins in total.

\subsubsection{The Correlation Functions Estimators}

Given the computed pair-counts from our fast correlation function estimator code, we use the~\citep{land93} estimator:
 \begin{equation}
\omega(\theta)=\frac{N_{dd}-2N_{dr}+N_{rr}}{N_{rr}}
\label{AngCorrEst}
\end{equation}
where $N_{dd}$ is the normalized number of data-data pairs counted within a certain angular separation., $\theta \pm \delta\theta$, over all SDSS DR7 fields. $N_{dr}$ and $N_{rr}$ stand for the number of data-random pairs and random-random pairs respectively. The data in the estimator can either stand for a galaxy or a galaxy pair. In all cases, we use approximately ten times as many randoms as data.

\begin{figure*}
\begin{center}
{{\includegraphics[width=1.\textwidth]{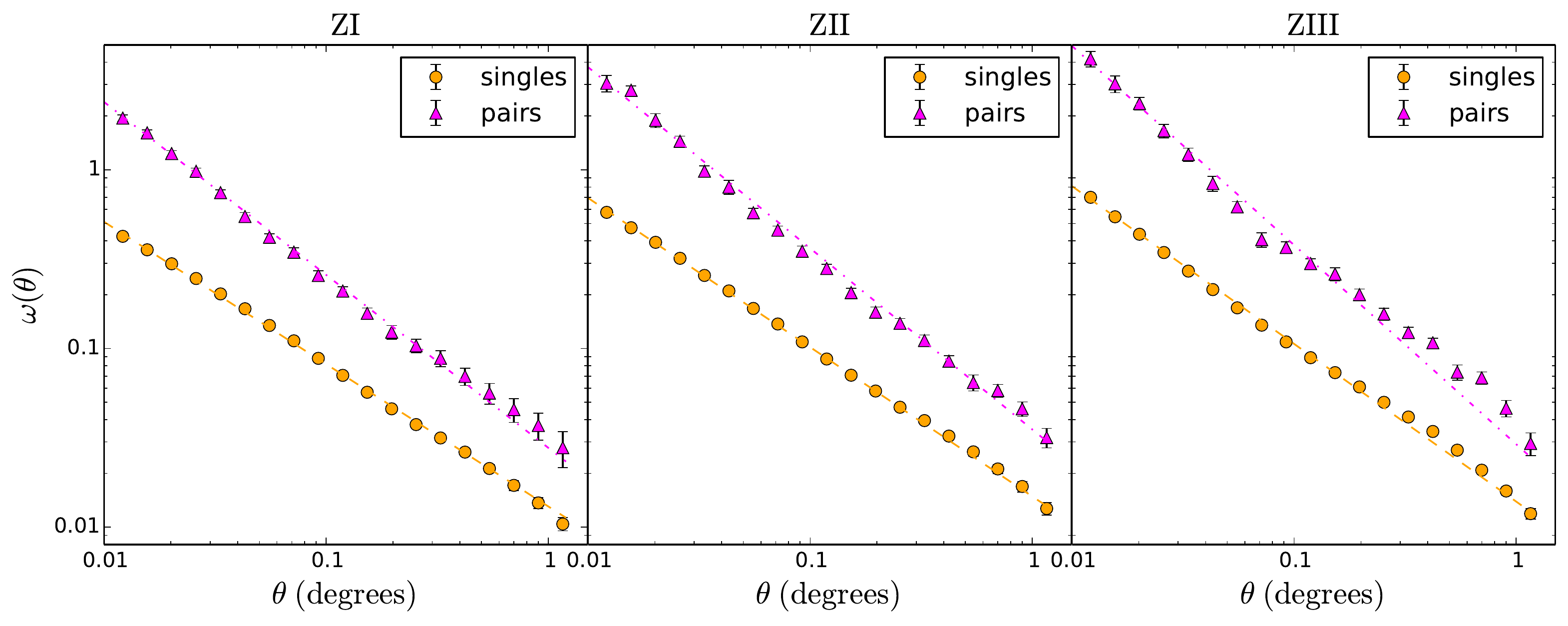}}}
\end{center}
\caption{From left to right: The two-point angular correlation function for all individual galaxies and galaxy pairs selected from the samples ZI, ZII, and ZIII. The best fit power-law models are shown by the dashed lines.
\label{singlePairs}}
\end{figure*}

\subsubsection{Errors and Covariance Matrices}

We calculate error bars by using the \textit{delete one jackknife} method, and we use a jackknife defined covariance matrix to fit the measured correlation functions. We determine the covariance matrix for N jackknife samples by using
\begin{equation}
C_{jk}(x_i, x_j) = \frac{N - 1}{N} \sum_{k=1}^{N} (x_i^k - \Bar{x_i})(x_j^k - \Bar{x_j})
\label{covmat}
\end{equation}
where $x_i$ is the $i^{th}$ measure of the statistic of N total samples, and the mean expectation of $x$ is given by:
\begin{equation}
\Bar{x_i} = \sum_{k=1}^{N} \frac{x_i^k}{N}
\end{equation}

To determine the best fit model for each measured correlation function, we perform a chi-squared minimization:
\begin{equation}
\chi^2 = \frac{1}{N_{dof}} \sum_{i,j} [\omega(\theta_i) - \omega_m(\theta_i)] C_{i,j}^{-1} [\omega(\theta_j) - \omega_m(\theta_j)],
\label{curfit}
\end{equation}
where $\omega(\theta)$ is the measured two-point galaxy angular correlation function, $\omega_m(\theta)$ is the model two-point galaxy angular correlation function, and $C_{i,j}$ are the elements of the covariance matrix calculated by using Equation~\ref{covmat}.

\subsection{The Angular Correlation Functions of Galaxies and Isolated Galaxy Pairs}\label{singlePairCorr}

We have previously~\citep{Wang13} demonstrated that the angular correlation function for galaxies drawn from the SDSS DR7 sample can be approximated with a power-law. In the three panels of Figure~\ref{singlePairs}, we now present the angular correlation functions measured for the individual galaxies and isolated galaxy pairs that are selected from the ZI, ZII, and ZIII samples (as defined in Section~\ref{volData}). As expected from hierarchical clustering models, the galaxy pairs are more strongly clustered than individual galaxies. In addition, we clearly see structure in the data that are traditionally interpreted as the transition from a one-halo to the two-halo term in the correlation function, but in this case for a pair of galaxies. 

Before proceeding with a more detailed analysis of the halo model interpretation of the galaxy pair angular correlation function, we first perform power law fits to all correlation functions by minimizing the $\chi^2$-statistic that is described in Equation~\ref{curfit}. The fit parameters: amplitude ($A_{\omega}$),  and slope ($\gamma$) are shown in Table~\ref{powerLawTable}. In Section~\ref{hod_singlePairs}, we will compare these fit parameters to the results generated by using our new HOD model.

\begin{table}
\begin{center}
\caption{The power-law fit parameters to the angular correlation function of galaxies and galaxy pairs selected from the ZI, ZII, and ZIII sub-samples.}
{\begin{tabular}{c  c  c  c c}
\hline
& & $log_{10}(A_{\omega})$ & $\gamma$ & $\chi_{pl}^2 / dof$ \\ \hline
\multirow{2}{*}{ZI} & Galaxies & -1.883$\pm$0.019 & 1.797 & 15.03\\
& Pairs &  -1.556$\pm$0.016 & 1.968 & 8.33 \\ \hline
\multirow{2}{*}{ZII} & Galaxies & -1.830$\pm$0.017 & 1.835 & 7.29 \\
& Pairs &  -1.453$\pm$0.015 & 2.014 & 5.42 \\ \hline
\multirow{2}{*}{ZII} & Galaxies & -1.857$\pm$0.018 & 1.8827 & 19.4 \\
& Pairs &  -1.536$\pm$0.015 & 2.1146 & 10.29 \\ \hline
\end{tabular}}
\label{powerLawTable}
\end{center}
\end{table}

\section{HOD Model Framework}\label{hod}

A halo occupation distribution (HOD) model quantifies the probability distribution for the number of galaxies, N, hosted by a dark matter halo as a function of the halo mass. The galaxy HOD model separates the two-point clustering signal of galaxies into two components: the distribution of galaxies within individual haloes, which dominates small scale clustering, and the clustering of galaxies between two different haloes, which dominates large scale clustering. By combining these two components, an HOD model can accurately model the observed scale-dependent clustering signal. {Throughout this paper, we assume a flat cosmology with $\Omega_m$ = 0.27, $h$ = 0.7, $\sigma_8$ = 0.8, and $\Gamma$ = 0.15.}

More formally, we use the halo model of galaxy clustering to generate a model spatial correlation function, $\xi(r)$, and subsequently project this three-dimensional spatial function to the two-dimensional angular sky coordinates. By comparing this projected angular clustering signal to the measured $\omega(\theta)$ from large area, photometric surveys, we can place constraints on the model parameters. The relationship between the angular and spatial correlation functions is provided via Limber's equation (assuming a flat Universe):
\begin{equation}
\omega(\theta) = \frac{2}{c} \int_{0}^{\infty} H(z)(dn/dz)^2 dz\int_0^{\infty} \xi(r = \sqrt{u^2 + x^2(z) \theta^2}) du
\end{equation}
The theoretical galaxy distributions, however, are more easily generated in Fourier space. Thus we need to convert theoretical power spectra into real-space correlations via the Fourier transform of the power spectrum:
\begin{equation}
\xi(r) = \frac{1}{2\pi^2}\int_{0}^{\infty}P(k,r)k^2 \frac{sin\ kr}{kr} dk
\end{equation}
In keeping with our HOD model, we split the theoretical galaxy power spectrum into two components: the contribution due to galaxies that reside within a single halo (the one-halo term), and the contribution due to galaxy pairs located in separate halos (the two-halo term):
\begin{equation}
P(k,r) = P_{1h}(k,r) + P_{2h}(k,r)
\end{equation}
In the rest of this section, we derive these theoretical  one-halo and two-halo terms, along with other important, associated quantities that are used in our HOD model. {In the following equations, we will use the notation $N(M) \equiv \ <N|M>$, $N_c(M) \equiv \ <N_c|M>$ and $N_s(M) \equiv \ <N_s|M>$, where $< N|M >$ is the probability distribution for the number of galaxies N hosted by a dark matter halo with mass M, and likewise for the associate probabilities for central, $<N_c|M>$, and satellite, $<N_s|M>$, galaxies.}

\subsection{The One-Halo Term}

To compute the power spectrum for the one-halo term, we recognize that we have two distinct components: the central-satellite galaxy component,  $P_{cs}(k,r)$, and the satellite-satellite component, $P_{ss}(k)$:
\begin{equation}
P_{1h}(k,r) = P_{cs}(k,r) + P_{ss}(k)
\end{equation}
Where r is the comoving separation between the galaxies. These individual theoretical power spectra terms, $P_{cs}$ and $P_{ss}$, are computed by using the following relations:
\begin{equation}
P_{cs}(k,r) = \int_{M_{vir}(r)}^{\infty} n(M)N_c(M)\frac{Ns(M)\mu(k|M)}{n_g^2/2} dM
\label{P1h_cs}
\end{equation}
\begin{equation}
P_{ss}(k) = \int_{M_{vir}(r)}^{\infty} n(M)N_c(M)\frac{(Ns(M)\mu(k|M))^2}{n_g^2} dM
\label{P1h_ss}
\end{equation}
$M_{vir}(r)$ is the minimum virial mass that can hold the corresponding central-satellite or satellite-satellite galaxy pairs at a separation $r$ for a given halo in Equation~\ref{P1h_cs} and~\ref{P1h_ss}:
\begin{equation}
M_{vir}(r) = \frac{4}{3}\pi r^3 \bar{\rho} \Delta
\end{equation}
In the previous equation, the comoving background density of the Universe is defined as $\bar{\rho}$ = 2.78 $\times$ $10^{11}\Omega_m\ h^2 M_{\sun} Mpc^{-3}$. We follow~\citet{Blake} to define $\Delta$ = 200 as the critical over-density for virialization, thus we can express the virial radius in terms of the halo mass, $r_{vir} = {3M}/({\Delta\times 4\pi\bar{\rho}})$.

Other terms in  Equations~\ref{P1h_cs} and~\ref{P1h_ss} include $n(M)$, the halo mass function, which will be discussed in more detail in Section~\ref{haloMassFunction}; $N_c(M)$, the mean occupation for central galaxies at halo mass $M$; and $N_s(M)$, the mean number of satellite galaxies within a single halo. From these terms, we can compute the model galaxy number density:
\begin{equation}
n_{g,mod} = \int_{0}^{\infty} n(M)\ N(M)\ dM
\label{ng_mod}
\end{equation}
We can directly compare our model, $n_{g,mod}$, and observed galaxy number densities to place constraints on our specific HOD model parameters (see Section~\ref{model_basic} for more details). We can thus match the $n_{g,mod}$ with $n_{g,obs}$ to determine one HOD parameter, $M_{cut}$, for a given ($M_0$, $\alpha$, $\sigma_{cut}$) set.

Finally, $\mu(k|M)$ is the Fourier transform of the NFW~\citep{NFW} halo density profile, $\rho(r|M)$:
\begin{equation}
\rho(r|M) = \frac{\rho_s}{(r/r_s)(1 + r/r_s)^2}
\end{equation}
where $r_s$ is the scale radius of the halo. $\rho_s$ is the dark matter density at the scale radius:
\begin{equation}
\rho_s = \frac{M}{4\pi r_s^3}[ln(1+c)-\frac{c}{(1+c)}]^{-1}
\end{equation}
where $c$ is the $concentration\ parameter$ defined as $c=r_{vir}/r_s$. 
\citet{Bullock} and~\citet{Zehavi04} have shown that this term can be expressed as:
\begin{equation}
c(M,z) = \frac{11}{(1+z)}(\frac{M}{M_c})^{-0.13}
\end{equation}
where $M_c$ is a parameterized cutoff mass in units of $h^{-1}\ M_{\sun}$. For our assumed cosmology, and is quantified by $log_{10}(M_c)$ = 12.56.

\subsection{The Two-Halo Term}
The two-halo power spectrum can be computed as:
\begin{multline}
P_{2h}(k,r) = P_{m}(k)\ \times \\
\bigg[\int_{0}^{M_{lim}(r)} n(M)\ b(M,r)\ \frac{N(M)}{n_g^\prime} \mu(k|M) dM\bigg]^2
\label{P2h}
\end{multline}
where $P_m(k)$ is the non-linear matter power spectrum at the survey redshift and all other terms, other than $M_{lim}(r)$ and $b(M, r)$, are as defined before for the one-halo term. The mass limit, $M_{lim}(r)$, is calculated by employing the``$n^\prime_g$-matched'' approximation~\citep{Tinker05}, which considers \textit{halo exclusion} effects and matches the \textit{restricted galaxy number density}, $n^\prime_g(r)$, as a function of physical separation, $r$. We can compute $n^\prime_g(r)$ for our HOD model:
\begin{equation}
n_g^\prime(r) = \int_{0}^{M_{lim}(r)} n(M)\ N(M)\ dM
\label{ng_prime}
\end{equation}
On the other hand, we can also compute $n_g^\prime(r)$ by employing the halo exclusion:
\begin{multline}
n_g^{\prime\ 2}(r) = \int_{0}^{\infty} n(M_1) N(M_1) dM_1 \times \\
\int_0^{\infty} n(M_2) N(M_2) P(r, M_1, M_2) dM_2
\label{ng_prime_haloExclu}
\end{multline}
where $P(r, M_1, M_2)$ measures the probability of non-overlapping halos of masses $M_1$ and $M_2$ at separation $r$. 

By analyzing simulations, \citet{Tinker05} obtained $P(y) = 3 y^2 - 2y^3$ for 0 $< y <$ 1, $P(y) = 0$ for $y <$ 0, and $P(y) = 1$ for $y >$ 1, where $y$ is connected to the virial radii:
\begin{equation}
y(r) = \frac{x(r) - 0.8}{0.29},\ x(r) = \frac{r}{R_1 + R_2} 
\end{equation}
where $R_1$ and $R_2$ are the virial radii corresponding to masses $M_1$ and $M_2$. Given a galaxy density $n^\prime_g(r)$ computed from Equation~\ref{ng_prime_haloExclu}, we can increase $M_{lim}(r)$  in Equation~\ref{ng_prime} until we obtain the same galaxy density for our HOD model. Having determined this mass limit, we can subsequently use $M_{lim}(r)$ in Equation~\ref{P2h} to calculate the two-halo power spectrum term.

The last remaining undefined term in Equation~\ref{P2h},  $b(M,r)$, is the scale-dependent bias, which can be written in the following form~\citep{Tinker05}:
\begin{equation}
b(M,r)^2 = b^2(M)\frac{[1 + 1.17\xi_m(r)]^{1.49}}{[1+0.69\xi_m(r)]^{2.09}}
\end{equation}
where $\xi_m(r)$ is the non-linear, matter correlation function, and $b(M)$ is the \textit{halo bias function}  that quantifies the relative bias of a halo of mass M with respect to the overall dark matter distribution~\citep{Sheth, Tinker05}:
\begin{multline}
b(\nu) = 1 + \frac{1}{\delta_{sc}} \times \\
\left[ q\nu + s(q\nu)^{1-t} - \frac{q^{-1/2}}{1 + s(1-t)(1-\frac{t}{2})(q\nu)^{-t}}\right]
\end{multline}
where the three parameters are assigned the following values: $q = 0.707$, $s = 0.35$, and $t = 0.8$.

The model spatial correlation function can be calculated from the two-halo term by using the Fourier transform of the power spectra:
\begin{equation}
\xi^\prime_{2h}(r) = \frac{1}{2\pi^2}\int_{0}^{\infty}P_{2h}(k,r)k^2 \frac{sin\ kr}{kr} dk
\end{equation}
As shown by~\citet{Tinker05}, we can correct $\xi^\prime_{2h}(r)$ to the true spatial correlation function, $\xi_{2h}(r)$, via:
\begin{equation}
1 + \xi_{2h}(r) = [\frac{n_g^\prime(r)}{n_g}]^2 [1 + \xi_{2h}^\prime(r)]
\end{equation}
We note that this correction only modifies the $\xi_{2h}(r)$ measurement at small spatial separations, where the $\xi_{1h}(r)$ signal is dominated by the one-halo term. Therefore, this correction is a negligible correction ($< 1\%$) to our final HOD model correlation functions.

\subsection{Halo Mass Function}\label{haloMassFunction}

\begin{table*}
\begin{center}
\caption{The best-fit HOD model parameters for the galaxies and galaxy pairs selected from the three volume limited samples. All masses are in units of $M_{\sun}h^{-1}$.}
{\begin{tabular}{c  c  c  c c c c c c c}
\hline \hline
& & $\alpha$ & $log_{10}(M_0)$ & $\sigma_{cut}$ & $log_{10}(M_{cut})$ & $\chi^2 / dof$ & $log_{10}(M_{eff})$ & $b_g$ & $f_c$ \\ \hline
\multirow{2}{*}{ZI} & Galaxies & 1.16$\pm^{0.03}_{0.02}$ & 12.99$\pm^{0.029}_{0.025}$ & 0.19$\pm^{0.22}_{0.18}$ & 11.63 & 22.41 &13.40&1.13&0.76\\
                                       & Pairs &  1.54$\pm^{0.04}_{0.03}$ & 14.20$\pm0.03$ & 0.28$\pm^{0.13}_{0.10}$ & 13.18 & 9.31 &13.78&1.90&0.84\\ \hline
\multirow{2}{*}{ZII} & Galaxies & 1.18$\pm^{0.025}_{0.021}$ & 13.19$\pm^{0.02}_{0.018}$ & 0.10$\pm^{0.12}_{0.09}$ & 11.87 & 14.26 &13.45&1.21&0.76\\
                                       & Pairs &  1.73$\pm{0.03}$ & 14.41$\pm^{0.028}_{0.024}$ & 0.08$\pm^{0.13}_{0.07}$ & 13.36 & 13.02 &13.92&2.11&0.89\\ \hline
\multirow{2}{*}{ZIII} & Galaxies & 1.46$\pm^{0.04}_{0.03}$ & 13.59$\pm^{0.024}_{0.02}$ & 0.16$\pm^{0.13}_{0.11}$ & 12.37 & 26.19 &13.51&1.45&0.79\\
                                       & Pairs &  1.03$\pm^{0.028}_{0.031}$ & 14.86$\pm^{0.014}_{0.012}$ & 0.07$\pm^{0.12}_{0.06}$ & 13.87 & 7.64 &13.94&2.47&0.89\\ \hline
\end{tabular}}
\label{hodTable}
\end{center}
\end{table*}

The \textit{halo mass function}, $n(M)$, quantifies the number density of haloes as a function of mass $M$, originally described by~\citet{Press}:
\begin{equation}
n(M)dM = \frac{\bar{\rho}}{M} f(\nu) d\nu
\end{equation}
The mass function, $f(\nu)$, is typically modeled in the following form:
\begin{equation}
f(M) = \frac{1}{2 \nu}a_1 exp (- |ln \sigma^{-1} + a_2|^{a_3})
\end{equation}
where $a_1$ = 0.315, $a_2$ = 0.61, and $a_3$ = 3.8 as determined from simulations by~\cite{Tinker05}. In this form, $\sigma$ is defined by $\sigma(M,z)\equiv \delta_{sc} / \sqrt{\nu}$, which is the variance of the linear power spectrum within a spherical top hat that contains average mass M:
\begin{equation}
\sigma^2(M,z) = \frac{D^2(z)}{2 \pi^2}\int_0^{\infty} k^2 P_{lin}(k) W^2(kR) dk
\label{vlps}
\end{equation}
The functional terms in Equation~\ref{vlps} include $W(x) = (3/x^3) [sin\ x - x\ cos\ x]$; $R = (3\ M/(4\ \pi\ \bar{\rho}))^{1/3}$; $D(z)$, which is the linear growth factor at redshift $z$; and $P_{lin}(k)$, which is the linear power spectrum at redshift zero.

\subsection{Halo Occupation Distribution Model}\label{model_basic}
The number of galaxies that populate a halo of mass $M$ can be described as
\begin{equation}
N(M) = N_c(M) \times (1+N_s(M))
\label{N_M}
\end{equation}
where $N_c(M)$ is the mean occupations for central galaxies, and $N_s(M)$ is the mean occupation for satellite galaxies, {for which we assume a Poisson distribution~\citep{Kravtsov}.} The form of Equation~\ref{N_M} implies that a halo can only host a satellite galaxy if it already hosts a central galaxy. We follow~\citet{Ross} to model the HOD with a softened transition for both the central and satellite galaxies:
\begin{equation}
N_c(M) = 0.5\ [1 + erf(\frac{log_{10} (M/M_{cut})}{\sigma_{cut}})]
\label{NcM}
\end{equation}
\begin{equation}
N_s(M) = 0.5\ [1 + erf(\frac{log_{10} (M/M_{cut})}{\sigma_{cut}})] \times (\frac{M}{M_0})^\alpha
\label{NsM}
\end{equation}
{where $M_{cut}$ is the halo mass that can host a single, central galaxy, and $M_0$ is the halo mass where the halo starts to host satellite galaxies in addition to the central galaxy.} This model has four free parameters, one of which, $M_{cut}$, we can remove by matching the observed galaxy density, $n_g$, to the model-derived number density of galaxies by using Equation~\ref{ng_mod}. 

\begin{figure*}
\begin{center}
{{\includegraphics[width=1\textwidth]{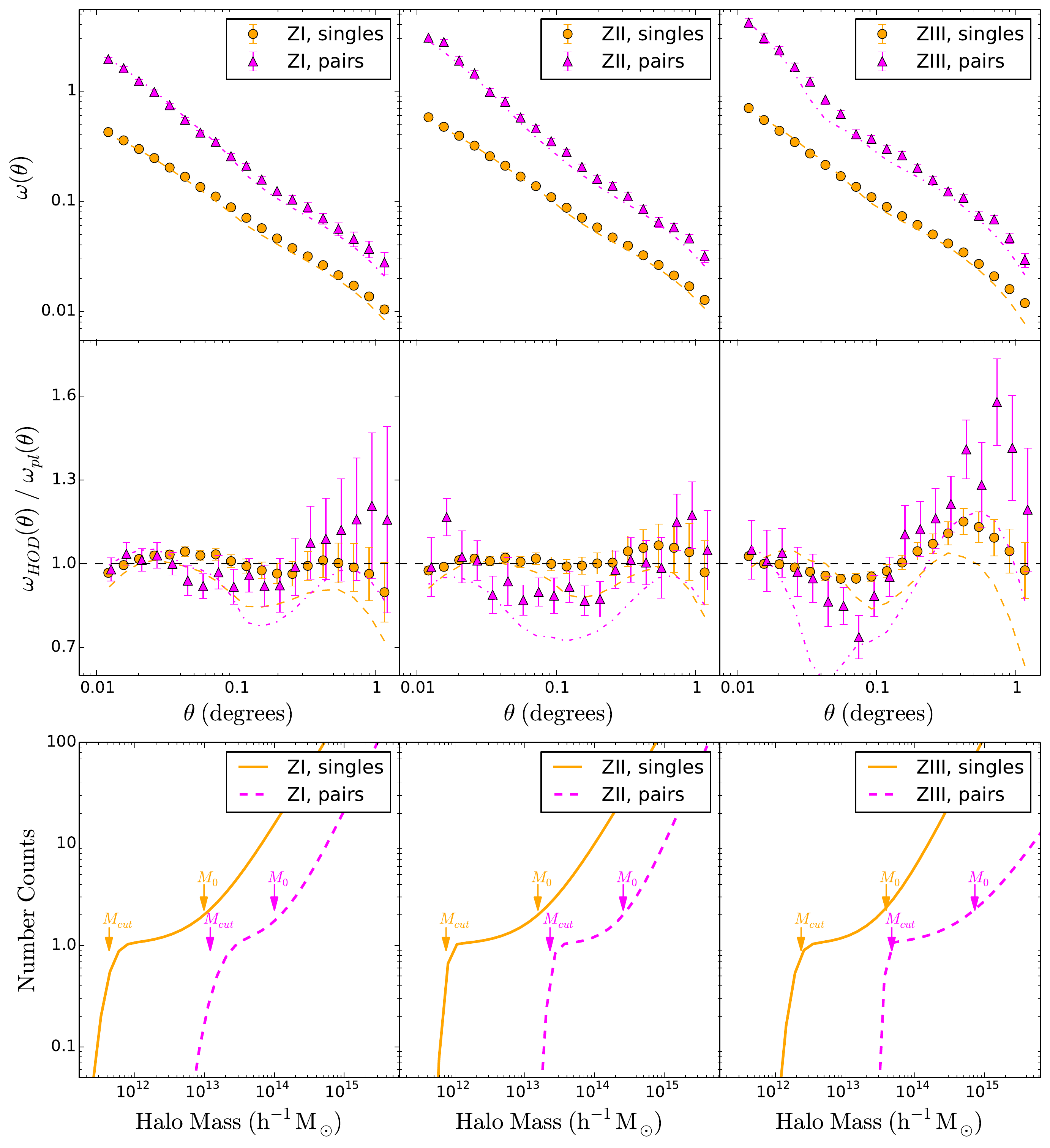}}}
\end{center}
\caption{The measured angular correlation function (top row) and HOD models for single galaxies (orange circles) and galaxy pairs (magenta triangles) selected from the ZI (left column), ZII (middle column) and ZIII (right column) samples. The dashed lines indicate the best-fit HOD models. The middle row displays the ratio of the best-fit HOD models (dashed line) and measured correlation function (points) to the best-fit power-law models. The bottom row displays the halo occupation distribution of the best-fit HOD models for the three single galaxy (orange) and galaxy pair (magenta) samples.}
\label{HOD_pairs}
\end{figure*}

To compare with previous results, we also derive two additional quantities from our HOD: the effective mass, $M_{eff}$: 
\begin{equation}
M_{eff} = \int M n(M) \frac{N(M)}{n_g} dM
\label{MeffEq}
\end{equation}
and the effective large-scale bias, $b_g$:
\begin{equation}
b_g = \int n(M) b(M) \frac{N(M)}{n_g} dM
\label{bgEq}
\end{equation}
We also use the halo mass function to weight the HOD to determine the average fraction of central galaxies in the sample:
\begin{equation}
f_c = \frac{\int n(M) N_c(M) dM}{\int n(M) N(M) dM}
\label{fcEq}
\end{equation}
The fraction for satellite galaxies is given by $f_s$ = $1- f_c$.

\begin{figure*}
\begin{center}
{{\includegraphics[width=1.\textwidth]{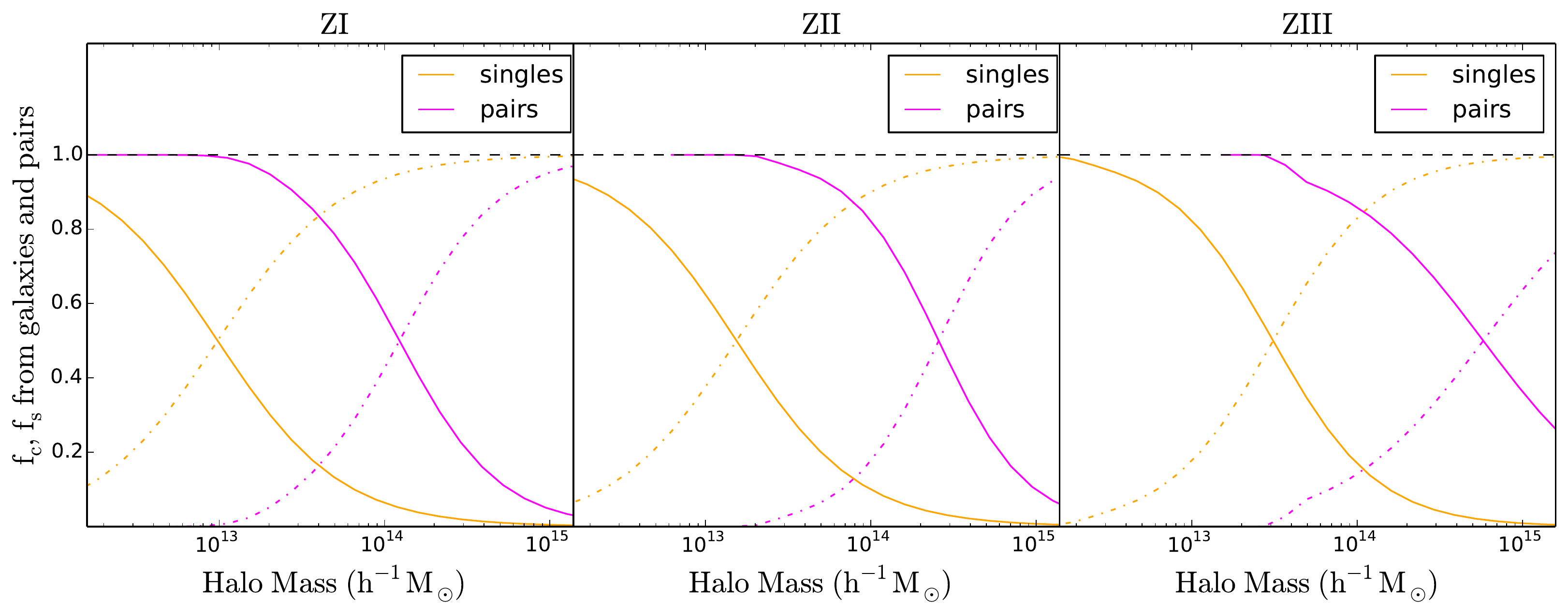}}}
\end{center}
\caption{The weighted fraction of central (solid lines) and satellite (dashed lines) components for the HOD models for single galaxies (orange) and galaxy pairs (magenta).}
\label{HOD_pairs_frac}
\end{figure*}

\section{HOD for Individual Galaxies and Galaxy Pairs}\label{hod_singlePairs}

We fit the 3-parameter halo model ($\alpha$, $M_0$, $\sigma_{cut}$) to the observed angular correlation functions, fixing the remaining parameter $M_{cut}$ by matching the model galaxy number density to the observed galaxy number density. By comparing the change in these parameters between different galaxy or galaxy pair samples, we can quantify the dependence of the clustering of galaxies on these parameters. We determine all HOD parameters by using a Monte Carlo Markov-Chain (MCMC) sampler to locate the minimum value of the $\chi^2$ fitting statistic by using the information from the full covariance matrix. Specifically, we use the \textbf{emcee}  code~\citep{Mackey} for which we set $nwalkers$ to 288, and we use at least 100 steps within each walk, which we have found to be sufficient for convergence. 

We use the peak probability as the location of the optimal fitting. Nominally this is given by the value where the $\chi^2$ measured by the fitting process is minimized. However, the MCMC approach can often give similar $\chi^2$ values at slightly different parameter combinations as we sample near the peak probability. We have verified that these different parameter configurations do not significantly change the results presented in the rest of this paper. We determine the errors on the fit parameters by finding the one sigma range around the peak probability. The best-fit HOD values and the standard deviations of each set of model parameters are listed in Table~\ref{hodTable}. The number of galaxy triplets and quads is too small within our samples to obtain correlation functions with sufficiently small error bars to make model constraints. Therefore, we only perform HOD fittings to the single galaxy and galaxy pair samples in the rest of this paper.

{Our photometric sample could potentially be affected by projection effects as the volumes probed by the angular cones increases. Furthermore, when working with galaxy pairs, we might expect a second projection effect to result when two, distinct galaxy pairs appear too close to reside in the same physical halo. For example, we might consider imposing a minimum angular separation of at least $0.028^\circ$ or $0.014^\circ$ at $z = 0.05$ or $0.1$ to force galaxy pairs into two, distinct massive halos. To test the effects of these projection concerns, we measure the correlation functions for the galaxy pairs between $0.01^{\circ}$ and $1^{\circ}$, and we also measure these same correlation functions between $0.02^{\circ}$ and $0.8^{\circ}$, which equates to removing the three smallest and two largest angular bins from the full angular range probed by the first set of correlation functions. We compared the HOD model fits to these two sets of correlation functions and found minimal variation between the measured HOD model parameters, in all cases the changes were less than one sigma, and were generally much less than this. Thus, for simplicity we simply used the full angular range for all correlation measurements of both the single galaxies and the isolated galaxy pairs.}

\subsection{Model Fits for the Three Samples}\label{3hod}

We fit the halo model to galaxy and galaxy pair samples selected in three volume-limited samples: ZI ($0.0 < z \leq 0.3$, $M_{r}\ < -19.0$), ZII ($0.0 < z \leq 0.3$, $M_{r}\ < -19.5$), and ZIII ($0.3 < z \leq 0.4$, $M_{r}\ < -19.5$). The full set of HOD fit parameters for these six samples are presented in Table~\ref{hodTable}, and the data and the best fit HOD models are shown in Figure~\ref{HOD_pairs}.

We display in the top panel of Figure~\ref{HOD_pairs} the measured $\omega(\theta)$ for galaxies and galaxy pairs that lie within the ZI, ZII and ZIII samples. The HOD model fits to all single galaxy correlation functions are computed between $0.01^{\circ}$ and $1^{\circ}$ (i.e., sixteen degrees of freedom), and yield $\chi^2/dof$ values ranging from 14.26 to 26.19. The model fits to the galaxy pairs are generally more accurate than for the single galaxies for all three samples; however, since these correlation functions tend to have larger error bars, this result is not surprising. With sixteen degrees of freedom, we have $\chi^2/dof$ values ranging from 7.64 to 13.02; and no systematic discrepancies are found with any of the model fits, implying that there are no major biases present in the model fits. 

We also note that the $\chi^2/dof$ values for the HOD model fits are not necessarily smaller than the $\chi^2_{pl}/dof$ values for the best power-law fits, which are tabulated in Table~\ref{hodTable}. The error bars on the correlation functions are extremely small for most angular bins, which will result in large $\chi^2$ fitting values even when there are only small deviations between the model and the data.

To highlight the differences between our best fit HOD and the power-law models to the measured correlation functions, we display the ratio of the best fit HOD model (dashed line) and measured correlation function (points) to the best fit power-law model in the middle panels of Figure~\ref{HOD_pairs}. The deviations from unity within these panels clearly indicate that the HOD models successfully reproduce the observed deviations from the power-law model and that the shape of the one-halo term (small scales) and the two-halo term (large scales) are closer or almost identical to the observed $\omega(\theta)$. Furthermore, we see that the HOD models capture the variation shown in the transition regions between the one-halo and two-halo terms for both single galaxies and galaxy pairs, which is missed by the power-law model. 

In the bottom panels of Figure~\ref{HOD_pairs}, we display the best-fit halo occupation distributions for individual galaxies and galaxy pairs in the three volume-limited samples, assuming the galaxy populations defined by Equations~\ref{N_M}.  The HOD for individual galaxies shows inflection points around $M_{cut} = 10^{11.63} h^{-1} M_{\sun}$ for the ZI sample, $M_{cut} = 10^{11.87} h^{-1} M_{\sun}$ for the ZII sample and around $M_{cut} = 10^{12.37} h^{-1} M_{\sun}$ for the ZIII sample;  this parameter defines the mass scale at which haloes host central galaxies. Likewise, these model fits define $M_{0} = 10^{12.99} h^{-1} M_{\sun}$, $M_{0} = 10^{13.19} h^{-1} M_{\sun}$, and $M_{0} = 10^{13.59} h^{-1} M_{\sun}$ for the ZI, ZII and ZIII samples respectively, which is the mass scale at which halos start to host satellite galaxies in addition to a central galaxy. Similarly, we find that the best-fit galaxy pair parameters are $M_{cut} = 10^{13.18} h^{-1} M_{\sun}$ and $M_{0} = 10^{14.20} h^{-1} M_{\sun}$ for the ZI sample, $M_{cut} = 10^{13.36} h^{-1} M_{\sun}$ and $M_{0} = 10^{14.41} h^{-1} M_{\sun}$ for the ZII sample and $M_{cut} = 10^{13.87} h^{-1} M_{\sun}$ and $M_{0} = 10^{14.86} h^{-1} M_{\sun}$ for the ZIII sample. 

\begin{figure*}
\begin{center}
{{\includegraphics[width=.67\textwidth]{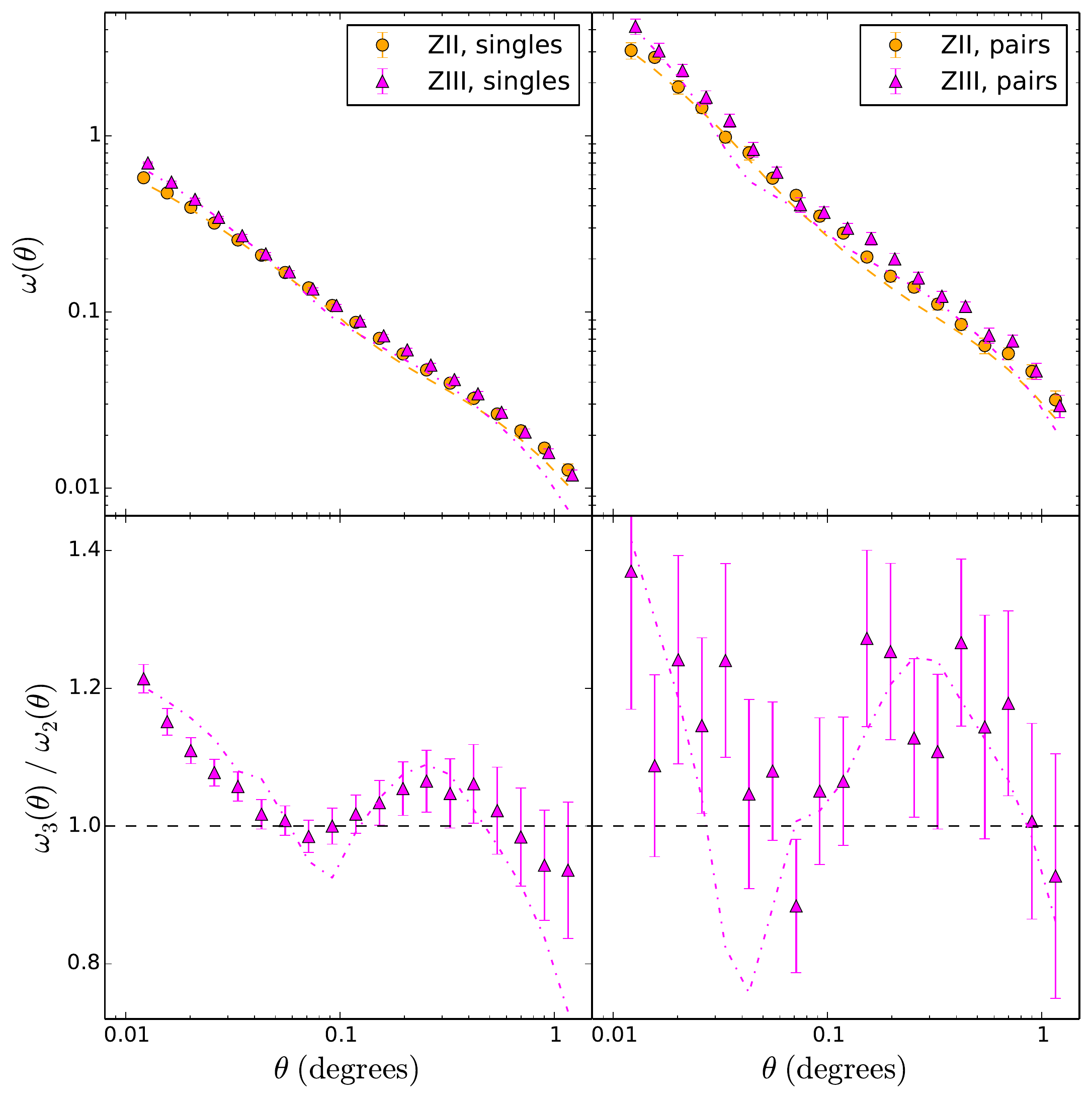}}}
\end{center}
\caption{The measured angular correlation function (top row) and HOD models for single galaxies (left column) and galaxy pairs (right column) that are selected from ZII sample (orange circles) and ZIII sample (magenta triangles). The dashed lines indicate the best-fit HOD models. The bottom row displays the ratio of the best-fit HOD models (dashed line) and measured correlation function (points) of the ZIII sample to the ZII sample.}
\label{HOD_evo}
\end{figure*}

From the figures and the tabulated fit parameters, we see that the $M_{cut}$ parameter for galaxy pairs is close to the value at which a halo hosts two galaxies as measured from the single galaxy correlation function. Naively, one might think the mass scale at which haloes host one central galaxy pair should be the same as the mass scale at which haloes host one central and one satellite galaxy. However, since we define our galaxy pair catalog as $isolated$ galaxy pairs (i.e., no nearby galaxies within a distance that is three times the group radius), we have parent haloes that are unlikely to reside in an environment that contains both small and large haloes. Therefore, our catalog preferentially selects more massive dark matter haloes that have higher probabilities to form $isolated$ pairs. Thus the haloes for our pair catalog that contain one central galaxy pair (two galaxies, i.e., where the $N(M) = 1$, usually at the point $\gtrapprox M_{cut}$) must be larger than the $M_0$ for the single galaxies. The average number of single galaxies hosted by the halo mass $M$=$10^{13}h^{-1}M_{\sun}$ is 2.0, 1.5, and 1.2 in the ZI, ZII and ZIII samples, respectively, and the average number of  galaxy pairs hosted by the halo mass $M$=$10^{14.5}h^{-1}M_{\sun}$ is likewise 4.9, 2.4 and 1.5.

\begin{table*}
\begin{center}
\caption{The best-fit HOD model parameters for galaxy pairs selected from the ZI sample as early-early pairs (\textit{EE}) or late-late pairs (\textit{LL}). All masses are in units $M_{\sun}h^{-1}$.}
{\begin{tabular}{c c c c c c c c c c c c}
\hline \hline
& $\alpha$ & $log_{10}(M_0)$ & $\sigma_{cut}$ & $log_{10}(M_{cut})$ & $\chi^2 / dof$ & $log_{10}(M_{eff})$ & $b_g$ &$f_c$& $log_{10}(A_{\omega})$ & $\gamma$ & $\chi_{pl}^2 / dof$ \\ \hline
EE & 1.40$\pm^{0.05}_{0.04}$ & 14.32$\pm^{0.04}_{0.03}$ & 0.30$\pm$0.17 & 13.42 &6.82&13.97&2.09&0.83&-1.66$\pm$0.01&2.09$\pm$0.02&2.88\\
LL &  0.77$\pm^{0.028}_{0.03}$ & 15.48$\pm^{0.04}_{0.05}$ & 1.17$\pm^{0.17}_{0.13}$ & 14.76 &1.27&13.79&1.77&0.99&-1.69$\pm$0.02&1.67$\pm$0.01&1.73\\ \hline
\end{tabular}}
\label{hodtypeTable}
\end{center}
\end{table*}

We compare the weighted fraction of central (given by Equation~\ref{fcEq}) and satellite galaxies and galaxy pairs as a function of the dark matter halo mass for all three galaxy samples in Figure~\ref{HOD_pairs_frac}.  In all samples, we see that the transition from central to satellite occurs at higher masses for galaxy pairs. This observation agrees with the standard picture that galaxy pairs are formed  in the cores of more massive dark matter haloes than single galaxies. We also note that since our group selection criteria selects \textit{isolated} galaxy pairs, we expect that there are a large fraction of galaxy pairs that are the only galaxies that populate a parent dark matter halo. Furthermore, if these galaxy pairs are \textit{real} physical pairs at close separation, we will only see central galaxy pairs accompanied by satellite galaxy pairs in the most massive haloes. At these mass scales, the haloes are large enough to host two galaxy pairs, while still allowing the two pairs to remain separated at a large physical range within the halo. Thus, the central and satellite galaxy pairs reside in subhaloes of the parent halo.

In Table~\ref{hodTable} we list the derived values of the effective halo mass, $M_{eff}$, and the galaxy bias factor, $b_g$, that are calculated by using Equations~\ref{MeffEq} and~\ref{bgEq}. The effective mass and the linear bias are $M_{eff}$ = $(10^{13.40},\ 10^{13.45},\ 10^{13.51})\ h^{-1}M_{\sun}$ and $b_g = $ (1.13, 1.21, 1.45) for the ZI, ZII, and ZIII samples, respectively. These results imply that the linear bias increases with luminosity and redshift, which agrees with fits to the large-scale power spectrum of the main SDSS DR7 galaxy samples~\citep{Hayes}. We compare our results in more detail with previous work in Section~\ref{results}. 

The effective halo mass and the linear bias for the galaxy pairs are $M_{eff}$ = $(10^{13.78},\ 10^{13.92},\ 10^{13.94})\ h^{-1}M_{\sun}$ and $b_g = $ (1.90, 2.11, 2.47), which, in comparison with our values for single galaxies, confirms that galaxy pairs are hosted by massive dark matter haloes and are highly biased tracers of the underlying dark matter distribution. The difference of $M_{eff}$ between the single galaxies and isolated galaxy pairs is consistently around $10^{\sim0.4}$, or 2.5. This ratio implies that the haloes hosting isolated galaxy pairs are, on average, slightly greater than twice the mass of the haloes that host single galaxies. We note that the systematic increase in galaxy bias between the ZI and the ZII and ZIII samples indicates that the bias is more dependent on sample redshift than on luminosity, at least within the redshift range probed by our volume limited samples. In addition, the similarity in $M_{eff}$ and the change in $b_g$ between the ZII and ZIII samples suggests that galaxy pairs in the higher redshift sample are more strongly clustered than the galaxy pairs in the lower redshift sample and that the parent dark matter haloes for the ZIII pairs are likely reside in more overdense environments.

Finally, we note that the fitting parameter, $\alpha$, is systematically larger for galaxy pairs than for the single galaxies in the two lower redshift samples:  ZI and ZII. $\alpha$ is the slope of the halo occupation distribution at increasing mass, which controls the rate of increase of satellite galaxies in a halo. This trend implies that isolated galaxy pairs at low redshift are more likely to be strongly clustered, and are probably embedded within even more massive dark matter haloes than single galaxies. On the other hand, the highest redshift sample, ZIII, has a shallower slope (or smaller value of $\alpha$), than the single galaxy sample. This implies that the largest haloes in this sample are more likely to be composed of single galaxies than isolated galaxy pairs. Thus, over time we see an increase in the clustering of galaxy pairs in the same mass haloes, which is consistent with the expected assembly history of galaxies and haloes.

\subsection{Redshift Dependence}\label{hodevo}

By comparing model fits to the correlation functions in the ZII ($0.0 < z \leq 0.3$, $M_{r}\ < -19.5$) and ZIII ($0.3 < z \leq 0.4$, $M_{r}\ < -19.5$) samples, we can study the redshift evolution of our HOD model (a complete analysis of the three samples is in Section~\ref{3hod}). We display in the top panels of Figure~\ref{HOD_evo} the measured $\omega(\theta)$ for galaxies and pairs from the ZII and ZIII samples, and we display the ratio of the $\omega(\theta)$ from two samples in the bottom panels. The ratios for the single galaxies and pairs show a similar trend, although the ratio for pairs shows more fluctuations that are due to the larger error bars in the pair correlations. Since the relative bias between two populations is defined by: $b^2_{1,2} = \omega_1/\omega_2$, we see that the deviations from unity indicate bias evolution between the ZII and ZIII samples. We also see scale dependence in the relative bias, as $b(\theta)$ changes with scale, becoming largest within the one and two halo regions, while being similar within the transition region between these two terms.

\subsection{Type Dependence\label{hodtype}}

We now focus our investigation on the dependence of the clustering of galaxy pairs on galaxy type. In order to ensure sufficient statistics, we only use the ZI sample, which we subdivide into the following two subsamples:
\begin{description}
\item \textbf{EE}: a galaxy pair in which both galaxies are identified as $early$-type galaxies, and
\item \textbf{LL}: a galaxy pair in which both galaxies are identified as $late$-type galaxies.
\end{description}
The galaxy type is determined by using the $m_u - m_r$ color cut, where $m_u$ and $m_r$ are dereddened model magnitudes. Following~\citet{Strateva}, if this color is $> 2.2$, the galaxy is classified as an early-type galaxy, while if it is $ < 2.2$ it is classified as a late-type galaxy.

\begin{figure}
\begin{center}
{{\includegraphics[width=.48\textwidth]{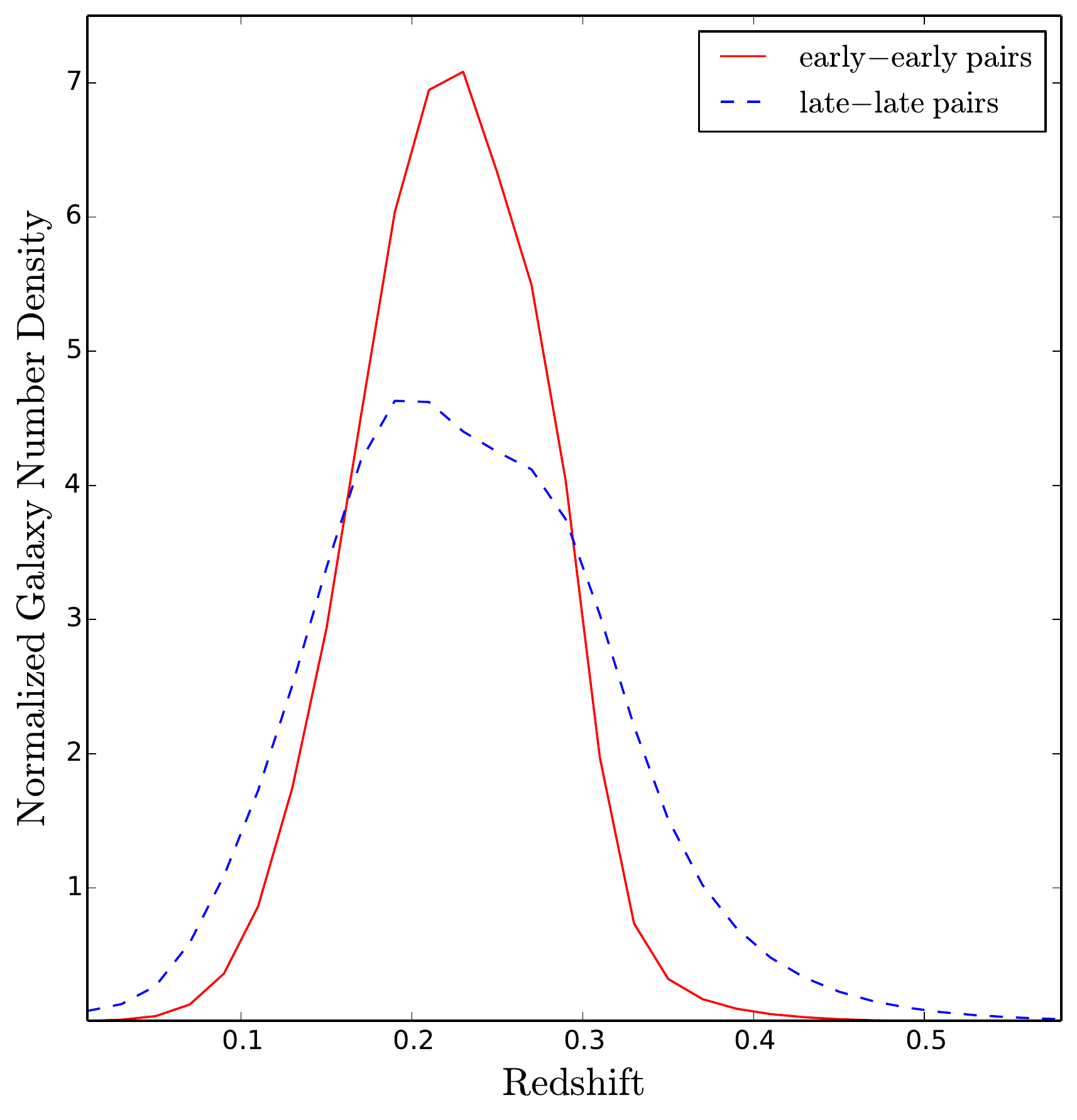}}}
\end{center}
\caption{The normalized number density distribution for early-early galaxy pairs (\textit{EE}, red), and late-late galaxy pairs (\textit{LL}, blue)}
\label{nz_pairs_types}
\end{figure}

\begin{figure}
\begin{center}
{{\includegraphics[width=.48\textwidth]{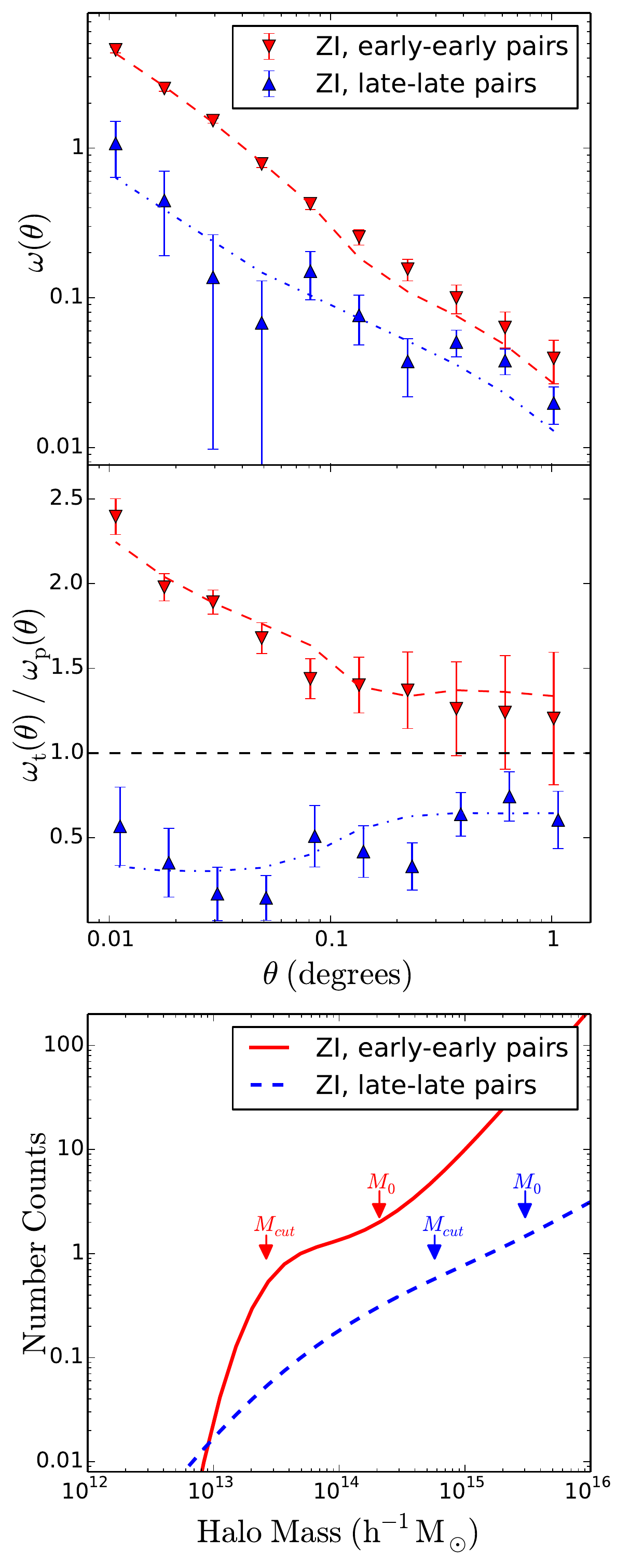}}}
\end{center}
\caption{The measured angular correlation function (top panel) and HOD models for galaxy pairs selected from ZI as early-early pairs (red up triangles) and late-late pairs (blue down triangles). The dashed lines indicate the best-fit HOD models. The middle panel displays the ratio of the best-fit HOD models (dashed line) and measured correlation function (points) to the parent sample. The bottom panel displays the halo occupation distribution of the best-fit HOD models for the two samples.}
\label{pair_types}
\end{figure}

These two sub-samples contain 68,316 and 17,291 isolated galaxy pairs, respectively. In Figure~\ref{nz_pairs_types} we present the normalized number density distributions for the two sub-samples as a function of redshift measured by using the technique discussed in Section~\ref{est_z}. To reduce the magnitude of the error bars on our correlation function measurements, we only use ten angular bins for all correlation function and associated covariance matrix measurements for these two sub-samples, and for the two sub-samples, \textit{L0.2} and \textit{L0.8} that will be discussed in section~\ref{hodlum}. However, to generate stable covariance matrices, we still use thirty-two jackknife samples.

We present the measured correlation function for the galaxies and galaxy pairs selected from the ZI sub-sample in the top panel of Figure~\ref{pair_types}. Also shown in these plots are the best-fit HOD models. The early-early galaxy pairs show stronger clustering over all scales than the late-late galaxy pairs, albeit with a steeper slope, in agreement with  galaxy morphology-density analyses~\citep[see, e.g.,][]{Dressler}.

These fit parameters are also listed in Table~\ref{hodtypeTable}. To quantify the dependence of these HOD model results for all galaxy pairs as a function of galaxy type, we divide both the measured correlation functions and the best-fit HOD models for our EE and LL subsamples by the best  fit HOD model for all galaxy pairs. The result is shown in the middle-panel of Figure~\ref{pair_types}, which, following the discussion on relative bias in Section~\ref{hodevo}, provides an indication of the square of the relative bias between these samples. The early-early type galaxy pairs show much stronger bias with respect to all galaxy pairs, especially at the halo center, while the late-late type galaxy pairs show less strong bias over all scales.

The bottom panel of Figure~\ref{pair_types} displays the best-fit halo occupation distributions for the two sub-samples. We find the late-late type galaxy pairs have larger $M_0$ and $M_{cut}$ values than the early-early type pairs, which is consistent with the results found for individual galaxies~\citep{Ross}. We note that the slope of the late-late galaxy pair HOD is smaller than the early-early galaxy pair HOD. Thus, the fraction of late-late galaxy pairs is largest in small mass haloes and smallest in high mass haloes.

\subsection{Luminosity Dependence}\label{hodlum}

We complete our analysis of the clustering of galaxies and galaxy pairs by examining the clustering dependence of galaxy pairs on galaxy luminosity. As presented in Section~\ref{hodtype}, to maintain sufficient statistics, we only use the ZI sample, which we subdivide into the following six subsamples:

\begin{figure}
\begin{center}
{{\includegraphics[width=.48\textwidth]{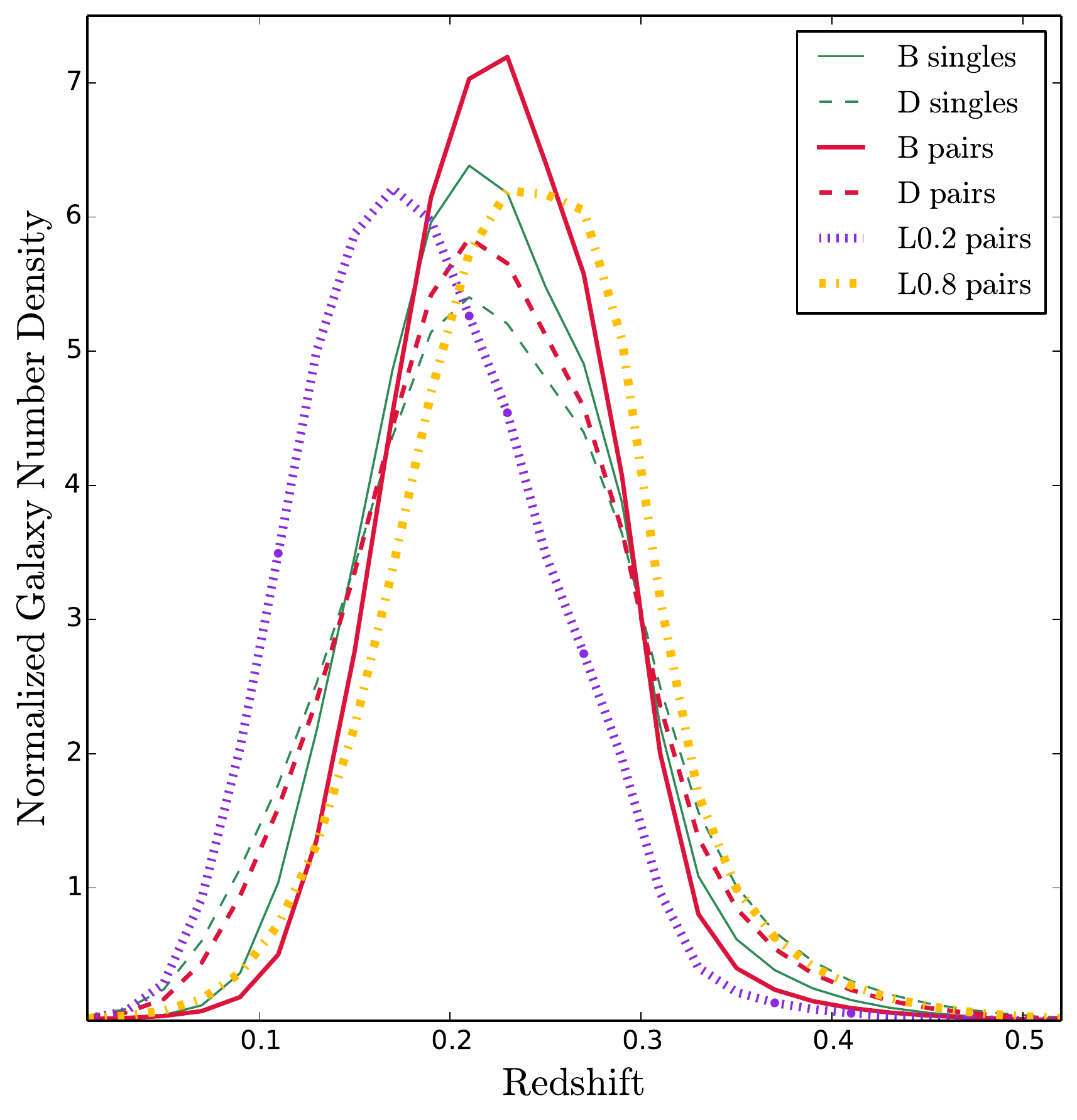}}}
\end{center}
\caption{The normalized number density distribution for bright single galaxies (\textit{B singles}), dim single galaxies (\textit{D singles}), bright galaxy pairs (\textit{B pairs}), dim galaxy pairs (\textit{D pairs}), large luminosity contrast pairs (\textit{L0.2}), and small luminosity contrast pairs (\textit{L0.8}).}
\label{nz_pairs_lums}
\end{figure}

\begin{description}
\item \textbf{B singles}: bright single galaxies with $M_{r}\ < -20.0$,
\item \textbf{D singles}: dim single galaxies with $-20.0\ <\ M_{r}\ < -19.0$,
\item \textbf{B  pairs}: bright galaxy pairs where both galaxies satisfy $M_{r}\ < -20.0$,
\item \textbf{D pairs}: dim galaxy pairs where both galaxies satisfy $-20.0\ <\ M_{r}\ < -19.0$,
\item \textbf{L0.2}: galaxy pairs where the luminosity ratio for the two galaxies is less than 20\%, and
\item \textbf{L0.8}: galaxy pairs where the luminosity ratio for the two galaxies is greater than 80\%.
\end{description}

\begin{table*}
\begin{center}
\caption{The best-fit HOD model parameters for bright single galaxies (\textit{B singles}), dim single galaxies (\textit{D singles}), bright galaxy pairs (\textit{B pairs}), dim galaxy pairs (\textit{D pairs}), large luminosity contrast pairs (\textit{L0.2}), and small luminosity contrast pairs (\textit{L0.8}). All masses are in units $M_{\sun}h^{-1}$.}
{\begin{tabular}{ l c c c c c c c c}
\hline \hline
& $\alpha$ & $log_{10}(M_0)$ & $\sigma_{cut}$ & $log_{10}(M_{cut})$ & $\chi^2 / dof$ & $log_{10}(M_{eff})$ & $b_g$ &$f_c$  \\ \hline
B singles & 1.3$\pm^{0.031}_{0.028}$& 13.43$\pm0.03$ & 0.21$\pm^{0.17}_{0.14}$ & 12.23 &4.04&13.42&1.34&0.77\\ 
D singles &  1.54$\pm^{0.039}_{0.042}$ & 13.46$\pm^{0.043}_{0.041}$ & 0.18$\pm^{0.17}_{0.15}$ & 11.81 &11.25&13.41&1.10&0.87\\ \hline 
B pairs & 0.96$\pm^{0.031}_{0.035}$ & 14.89$\pm^{0.038}_{0.042}$ & 0.63$\pm^{0.25}_{0.26}$ & 14.07 &1.85&14.06&2.17&0.93\\ 
D pairs &  1.41$\pm^{0.053}_{0.051}$ & 14.58$\pm^{0.052}_{0.047}$ & 0.34$\pm$0.16 & 13.46 &3.04&13.80&1.91&0.92\\ \hline 
L0.2 & 0.83$\pm^{0.044}_{0.49}$ & 15.40$\pm^{0.037}_{0.034}$ & 0.78$\pm^{0.35}_{0.26}$ & 14.15 &0.46&13.94&1.88&0.98\\ 
L0.8 &  1.83$\pm^{0.057}_{0.055}$ & 14.85$\pm^{0.049}_{0.53}$ & 0.001$\pm^{0.140}_{0.001}$ & 13.71 &2.10&14.11&2.38&0.95\\ \hline
\end{tabular}}
\label{hodlumTable}
\end{center}
\end{table*}

\begin{figure*}
\begin{center}
{{\includegraphics[width=1.\textwidth]{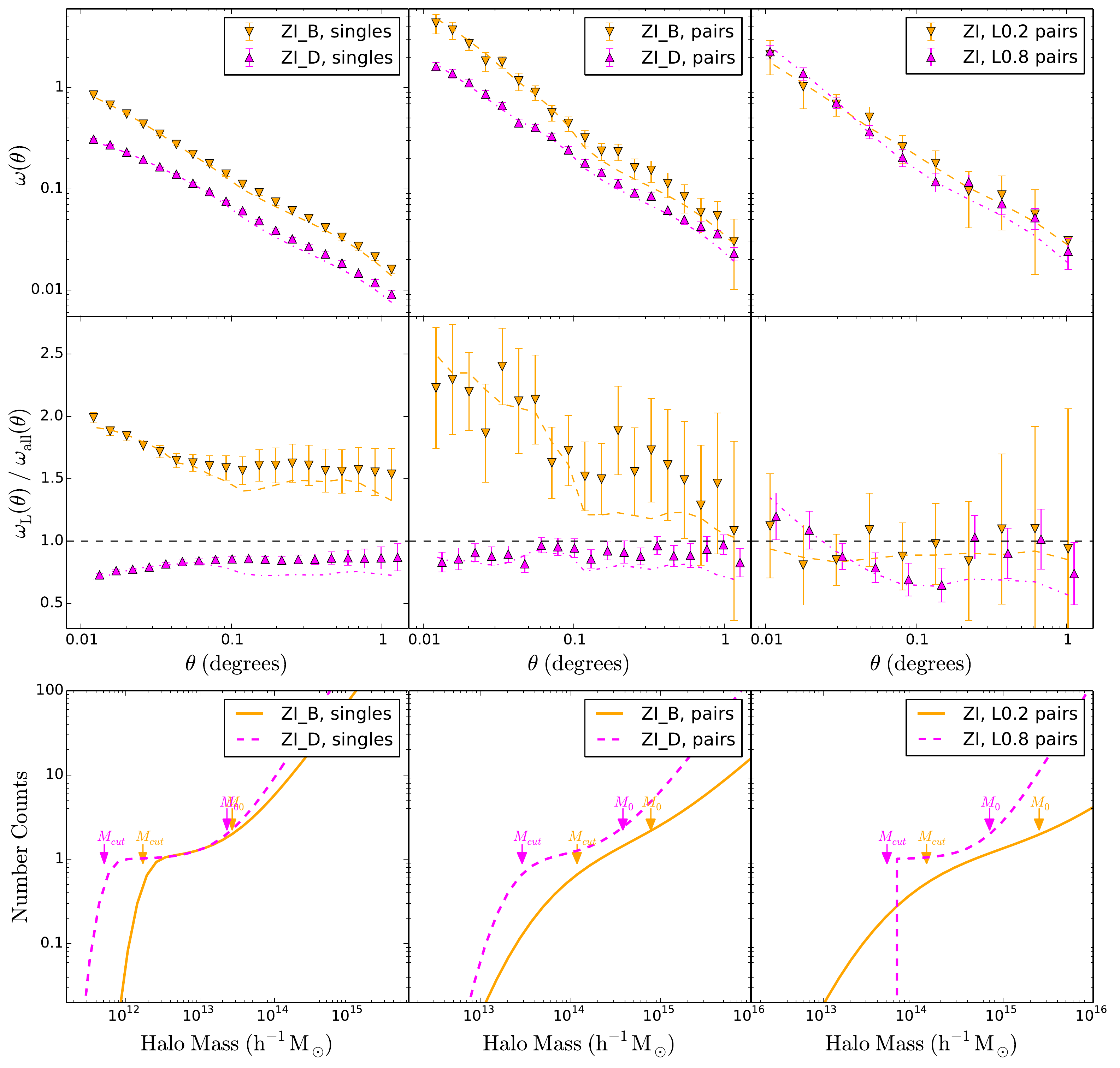}}}
\end{center}
\caption{The measured angular correlation function (top row) and HOD models for (left column) bright single galaxies (\textit{B singles}, orange down triangles) and dim single galaxies (\textit{D singles}, magenta up triangles) selected from the ZI sample, for (middle column) bright galaxy pairs selected from ZI (\textit{B pairs}, orange down triangles) and dim galaxy pairs (\textit{D pairs}, magenta up triangles), and for (right column) galaxy pairs selected from ZI as large luminosity contrast pairs (\textit{L0.2}, orange down triangles) and small luminosity contrast pairs (\textit{L0.8}, magenta up triangles). In all cases, the dashed lines indicate the best-fit HOD models. The middle row displays the ratio of the best-fit HOD models (dashed line) and measured correlation function (points) to the parent samples. The bottom row displays the halo occupation distribution of the best-fit HOD models for the two samples displayed in the first row.}
\label{pair_lums}
\end{figure*}

These six sub-samples contain 992,275, 2,983,258, 16,340, 67,153, 14,263, and 25,435 single galaxies or isolated galaxy pairs, respectively. In Figure~\ref{nz_pairs_lums} we present the normalized number density distributions for the six sub-samples as a function of redshift measured by using the technique discussed in Section~\ref{est_z}. As mentioned in Section~\ref{hodtype}, we only use ten angular bins for the sub-samples \textit{L0.2} and  \textit{L0.8} to reduce the magnitude of the error bars. For the other four sub-samples, we use nineteen angular bins; and in all cases, we still use thirty-two jackknife samples to generate the covariance matrices.

In the top-panel of Figure~\ref{pair_lums}, we present the measured correlation function for the six sub-samples as well as the best-fit HOD models. We first see that the bright samples for both single galaxies and galaxy pairs ($B\ singles$ and $B\ pairs$) show stronger clustering than their corresponding dim samples ($D\ singles$ and $D\ pairs$). On the other hand, the large and small luminosity contrast pairs demonstrate similar clustering, within the one $\sigma$ error bars, over all scales. The small luminosity contrast sample, however does have a local minimum near $0^{\circ}.1$, which we interpret as a result of the transition between the one-halo and two-halo terms.

The HOD model-fits for the six luminosity sub-samples are all well constrained with minimum $\chi^2$ ranging from 0.46 to 11.25 (sixteen degrees of freedom for the first four samples and seven degrees of freedom for the \textit{L0.2} and \textit{L0.8} samples). Together with these HOD model parameters, we list the effective halo mass and the galaxy bias factor for all sub-samples in Table~\ref{hodlumTable}. 

In the middle row of Figure~\ref{pair_lums}, we display the ratio of the measured correlation functions (points) and the best-fit HOD models (curves) to the same measurements from the parent ZI sample. As discussed previously, these ratios are the square of the relative bias between the two subsamples. Interestingly, In the left-panel we see that the bright single galaxies have a larger bias value, which increases toward the halo center, while the dim single galaxies show a lower bias at all scales, which decreases toward the halo center. This is not surprising if we recall that the parent sample contains galaxies of all magnitudes, so that we are seeing luminosity dependent bias.

In the middle-panel, we see the pairs follow a similar trend to their corresponding single galaxy sample; however we now see a stronger deviation in the data for the two-halo term. Despite larger error bars that can introduce uncertainties in the HOD model fitting, this trend is still seen within the $1\sigma$ deviations. We interpret this feature as an indication that bright galaxy pairs are more strongly clustered at scales larger than the halo size, thus they display a stronger bias on two-halo scales. In the right-panel of the middle row, we see that the relative bias of the galaxy pairs with large luminosity contrast (orange) to the parent sample is nearly uniform, while the galaxy pairs with similar luminosities (magenta) deviates both at the halo center and in the transition region.

The bottom panel of Figure~\ref{pair_lums} displays the best-fit halo occupation distributions for the six sub-samples. We find the bright samples have larger $M_0$ and $M_{cut}$ values than the dim samples. We note that the slope of all of the galaxy pairs' HODs are smaller than that of the single galaxies, thus these pairs are less likely to form larger haloes than single galaxies. Therefore, the fraction of galaxy pairs at the center of a halo is larger than fraction of single galaxies. The weighted fractions of the central galaxies and galaxy pairs from the six sub-samples (\textit{B\ singles}, \textit{D\ singles}, \textit{B\ pairs}, \textit{D\ pairs}, \textit{L0.2}, and \textit{L0.8}) are 0.77, 0.87, 0.93, 0.92, 0.98, and 0.95, respectively.

In Table~\ref{hodlumTable}, we present the HOD model fit parameters and the derived values of the effective halo mass, galaxy bias factor, and the fraction of central galaxies or galaxy pairs. Interestingly, we find that the \textit{L0.8} sub-sample has a much larger bias than the  \textit{L0.2} sub-sample, although they have similar measured angular correlation functions. In fact, the $b_g$ for the \textit{L0.2} sub-sample is similar to that of the ZI pair sample, which implies that large luminosity contrast pairs occupy similar dark matter haloes as normal pairs, while the large $b_g$ for the \textit{L0.8} sub-sample implies that pairs with similar luminosities occupy much larger dark matter haloes than the average ZI galaxy pair.

\section{Conclusions}\label{results}

We have measured the angular, two-point correlation function for galaxies drawn from three, volume limited samples of SDSS DR7: (ZI) $0.0 < z < 0.3,\ M_r < -19.0$, (ZII) $0.0 < z < 0.3,\ M_r < -19.5$, and (ZIII) $0.3 < z < 0.4,\ M_r  < -19.5$. We also have selected isolated galaxy pairs from these three volume-limited galaxy samples, and have measured the angular two-point correlation function for these galaxy pairs. By modeling the angular two-point correlation function, we have computed, for the first time, the best-fit halo model for photometrically selected isolated galaxy pairs. These models quantify that the galaxy pairs have larger effective mass and higher bias values than single galaxies. The central fraction for galaxy pairs is also higher than for single galaxies, which implies that galaxy pairs preferentially reside in dark matter haloes as central rather than satellite galaxies. 

Furthermore, we have explored the dependence of the correlation function and best-fit HOD models on redshift, galaxy type, and luminosity. We find that (1) early-early pair galaxy pairs have stronger clustering than late-late type galaxy pairs, (2) bright galaxy pairs have stronger clustering than dim galaxy pairs, and that (3) the clustering of large luminosity contrast pairs is similar (within the one-$\sigma$ error bars) to that of small luminosity contrast pairs.

We can directly compare our measured galaxy angular two-point correlation functions and associated best-fit HOD models to the results from~\citet{Ross}, who performed a similar best-fit HOD model analysis for galaxies selected from the SDSS DR5. The galaxy samples they use have similar luminosity and redshift cuts that approximate ours (e.g., their Z3 sample corresponds to our ZII sample). We also note that their sample shares a similar median redshift with our catalog (ZII: $\bar{z}\ \sim 0.21$, Z3: $\bar{z}\ \sim 0.25$), strengthening the comparison.

First, we note that the effective mass for single galaxies in their Z3 sample is $M_{eff} = 10^{13.11}\ h^{-1}M_{\sun}$, where we find a larger value of $M_{eff} = 10^{13.45}\ h^{-1}M_{\sun}$. The discrepancy is likely a result of the differences in the number densities of the two samples, our number density for single galaxies in ZII is smaller ($n_g \sim 0.0053\ h^3\ \rm Mpc^{-3}$) than the number density for their Z3 galaxies ($n_g \sim 0.0102\ h^3\ \rm Mpc^{-3}$). Thus, our catalog contains fewer low mass haloes. The smaller $n_g$ in our catalog is a result of the stronger requirements we developed to create our clean galaxy catalog that has minimal systematic effects~\citep{Wang13}. Second, the satellite galaxy fraction that we find for ZII, $f_s = 1 - f_c = 0.24$, is slightly higher than their results ($f_{s,Z3}$ = 0.15), but this value is in agreement  with other results~\citep[see, e.g, Figure 5][]{Zheng07}. We have luminosity thresholds of $L/L_* = 0.27$ for $M_r < -19.0$ and $L/L_* = 0.42$ for $M_r < -19.5$, which correspond to $f_s = $(0.29, 0.24) for single galaxies in our ZI and ZII samples.

On the other hand, we do not see the local minimum for the late-type single galaxies in our best-fit HOD model as seen previously~\citep{Zehavi11, Ross}. There are two reasons for this difference. First, we use different HOD models for the late-type galaxies. For example, \cite{Zehavi11} separate the central and satellite galaxies and measure the fractions of late-type galaxies seen in each case, while we simply apply the HOD model to the late-late galaxy pairs directly. Our HOD models for the late-late galaxy pairs are excellent fits, which implies that our model is sufficient enough to describe the distributions of late-late type galaxy pairs within dark matter haloes. Second, we only compute HOD fits to galaxy pairs within our catalog. The local minimum for the late-type single galaxies comes from low mass haloes, where the fraction of the single central late-type galaxy decreases as the mass increases. However, the halo mass is insufficient to host satellite galaxies as required to increase the halo occupation distribution. On the other hand, for the late-late type galaxy pairs, the parent haloes are massive enough to host additional galaxies, and they thus produce similar HOD models as the other sub-samples, which do not include an obvious inflection point.

Our clustering measurements of the early-early galaxy pairs show that they cluster more strongly than the late-late galaxy pairs. This is in agreement with the results from~\citep{HearinAgM, WatsonSFR} who studied the co-evolution of galaxies and halos. In their approach, they introduce a parameter, $z_{starve}$, that quantifies the epoch  in the halo's mass assembly history at which the star formation in the resident galaxies of the halo becomes inefficient. By implementing an age distribution match, they find $z_{starve}$, on general, is larger for redder (i.e., our early-type) galaxies than bluer (i.e., our late-type) galaxies, which implies that redder galaxies tend to reside in older haloes. As a result, we can expect that early-early galaxy pairs also preferentially reside in more massive haloes where the halo clustering strength is larger, which is what we observe.

We note this work only considered angularly selected isolated galaxies, which can suffer from line-of-sight contamination. Any interloper galaxies will systematically decrease the correlation functions that we use for the HOD modeling, and will, therefore, lower the final halo occupation distribution statistics. One technique to overcome this limitation would be to adopt accurate photometric redshift probability distribution functions~\citep[PDFs;][]{TPZ,SOMZ} to place stronger spatial constraints on galaxy pairs within our sample. The use of photometric redshift PDFs will allow for a more reliable determination of the evolution of galaxy pairs within dark matter haloes. 

{Finally, we note that we have cut our galaxy catalog into several subsamples: individual galaxies, galaxy pairs, and galaxies by type and luminosity. We have not, however, measured the cross-correlations between these subsamples. Such an investigation would be of direct physical interest as they will likely yield important insights into the clustering dependence of galaxy groups on richness, redshift, type, and luminosity. We intend to conduct a detailed study of these relationships once we have also more thoroughly addressed the projection effect issues.}

\section*{Acknowledgements}
The authors acknowledge support from the National Science Foundation Grant No. AST-1313415. RJB has been supported in part by the Institute for Advanced Computing Applications and Technologies faculty fellowship at the University of Illinois. This work also used resources from the Extreme Science and Engineering Discovery Environment (XSEDE), which is supported by National Science Foundation grant number OCI-1053575. We also thank Ashley Ross and Chris Blake for helpful discussions and comments that helped this work.

Funding for the SDSS and SDSS-II has been provided by the Alfred P. Sloan Foundation, the Participating Institutions, the National Science Foundation, the U.S. Department of Energy, the National Aeronautics and Space Administration, the Japanese Monbukagakusho, the Max Planck Society, and the Higher Education Funding Council for England. The SDSS Web Site is http://www.sdss.org/.

The Sloan Digital Sky Survey (SDSS) is a joint project of the american Museum of Natural History, Astrophysical Institute Potsdam, University of Basel, University of Cambridge, Case Western Reserve University, University of Chicago, Drexel University, Fermilab, the Institute for Advanced Study, the Japan Participation Group, Johns Hopkins University, the Joint Institute for Nuclear Astrophysics, the Kavli Institute for Particle Astrophysics and Cosmology, the Korean Scientist Group, the Chinese Academy of Sciences (LAMOST), Los Alamos National Laboratory, the Max-Planck-Institute for Astronomy (MPIA), the Max-Planck-Institute for Astrophysics (MPA), New Mexico State University, Ohio State University, University of Pittsburgh, University of Portsmouth, Princeton University, the United States Naval Observatory, the University of Washington. 

\bibliographystyle{mn2e}
\bibliography{2013_Wang_Group}

\bsp
\label{lastpage}

\end{document}